\documentclass[preprint,12pt]{elsarticle}
\journal{Thin-Walled Structures}
\usepackage{graphics,epsfig}
\usepackage{graphicx}
\usepackage{float}
\usepackage{amssymb,amstext,amsmath}
\usepackage{mathtools,physics}
\usepackage{booktabs}

\begin{document}
\begin{frontmatter}
\title{Asymptotically exact dimension reduction of functionally graded anisotropic rods}
\author{K.~C. Le$^{a,b}$\footnote{Phone: +84 093 4152458, email: lekhanhchau@tdtu.edu.vn}}
\address{$^a$Mechanics of Advanced Materials and Structures, Institute for Advanced Study in Technology, Ton Duc Thang University, Ho Chi Minh City, Vietnam
\\
$^b$Faculty of Civil Engineering, Ton Duc Thang University, Ho Chi Minh City, Vietnam
}
\begin{abstract} 
This study utilizes the variational-asymptotic method to establish a one-dimensional theory for functionally graded rods characterized by general anisotropy from the three-dimensional elasticity theory. A distinctive feature of this dimension reduction procedure is the numerical solution of dual cross-sectional problems, which provide rigorous upper and lower bounds for the average transverse energy density. By employing the Prager-Synge identity, we derive an error estimate in the energetic norm to establish the asymptotic exactness of the model. This estimate is extended to the dynamic regime for low-frequency vibrations. Furthermore, the dynamic validity of the theory is confirmed by comparing the one-dimensional dispersion relations with exact analytical three-dimensional solutions for wave propagation in composite rods. The results show that the developed one-dimensional model captures the long-wave asymptotic behavior of the three-dimensional elastic body with high fidelity. Numerical benchmarks indicate that while the naive rod theory incurs errors up to $20\%$ in deflection predictions, the current VAM framework reduces this discrepancy to below $3\%$, with log-log convergence studies confirming the theoretical $O(h/L)$ accuracy.
\end{abstract}


\end{frontmatter}

\section{Introduction}

Functionally graded (FG) materials \cite{niino1990recent,koizumi1992recent}, characterized by their continuous spatial variation of mechanical properties, have revolutionized the design of advanced structural components by allowing for the tailoring of performance criteria such as thermal resistance, weight reduction, and stress distribution \cite{saleh202030}. Despite their potential, the structural modeling of FG rods remains a formidable challenge. The inherent complexity of three-dimensional (3D) elasticity for these media--particularly when combined with curved geometries and initial twist--precludes analytical solutions except in highly idealized cases \cite{mian1998exact,horgan1999pressurized,sankar2001elasticity,pan2003exact,kashtalyan2004three,chu2015two}.

Traditional one-dimensional (1D) beam models, ranging from classical Euler-Bernoulli and Timoshenko theories \cite{sankar2001elasticity,aydogdu2007free,csimcsek2009free} to various refined higher-order theories \cite{li2008unified,thai2012bending,bourada2015new,reddy2021theories}, frequently rely on a priori kinematic assumptions. While these assumptions simplify the governing equations, they often fail to capture the subtle local effects caused by significant variations in elastic moduli and Poisson’s ratios across the cross-section. To address these limitations without sacrificing the efficiency of a 1D model, we employ the variational-asymptotic method (VAM) developed by Berdichevsky \cite{berdichevsky1979variational}. This approach, applied by himself to inhomogeneous and anisotropic rods \cite{berdichevskii1981energy}, allows for the systematic and rigorous dimensional reduction of the 3D action functional into a 1D theory that is asymptotically exact as the cross-sectional diameter $h$ approaches zero. The power of VAM lies in its ability to decouple the 3D elastic problem into a two-dimensional (2D) cross-sectional analysis and a 1D problem without resorting to ad hoc kinematic hypotheses. While Berdichevsky’s seminal work \cite{berdichevskii1981energy} laid the groundwork for inhomogeneous and anisotropic rods, the extension to FG materials with arbitrary anisotropy and material gradation requires a more nuanced numerical treatment of the transverse energy, especially when the Poisson's ratio varies over the cross section \cite{yu2004elasticity,yu2005generalized,le2020asymptotically,le2025asymptotically}.

Building upon our own works \cite{le2020asymptotically,le2025asymptotically}, we propose a comprehensive framework for FG rods. It is worth noting that the asymptotic treatment of fully anisotropic beams was significantly advanced by Volovoi et al. \cite{volovoi1999asymptotic}, who investigated the static behavior and end-effects of anisotropic I-beams. However, their analysis was restricted to homogeneous material properties and specific cross-sectional geometries. The present work extends the variational-asymptotic approach to accommodate the simultaneous effects of arbitrary anisotropy and continuous material gradation across the cross-section. We propose a framework where the cross-sectional warping functions are determined through dual variational problems. By solving these problems numerically, we obtain rigorous upper and lower bounds for the average transverse energy density, a technique that ensures the convergence of the effective 1D stiffness coefficients even for high-contrast material gradations. We also establish the method to restore the 3D stress and strain field from the solution of the 1D problem that has not been considered before. Furthermore, we provide a formal proof of asymptotic exactness in the energetic norm by utilizing the Prager-Synge identity, a methodology echoed in the work of Ladevèze and Simonds \cite{ladeveze1998new} for beam error estimation.

A significant contribution of this work is the extension of this asymptotic exactness to the dynamic regime. By reconciling the long-wave limit of the 3D Pochhammer-Chree type dispersion relations for layered anisotropic fibers \cite{nayfeh1996general} with our 1D model (see also \cite{armenakas1967propagation,chakraborty2003a}), we demonstrate that the proposed theory captures the essential physics of wave propagation with high fidelity.

The remainder of this paper is organized as follows. Section 2 establishes the variational framework for functionally graded curved and initially twisted rods. In Section 3, an asymptotic decomposition of the three-dimensional energy functional is performed, accounting for general anisotropy. Sections 4 and 5 develop the primal and dual cross-sectional variational problems, respectively, which provide the theoretical basis for the energy bounds. Section 6 describes the finite element implementation used to evaluate the effective 1D stiffnesses. The resulting one-dimensional theory and the proposed procedure for the reconstruction of the three-dimensional state are detailed in Section 7. Rigorous error bounds based on the Prager-Synge identity are established in Section 8, followed by a dynamic validation via dispersion analysis in Section 9. Section 10 offers concluding remarks.

\section{Variational formulation for FG-rods}
\begin{figure}[htb]
    \begin{center}
    \includegraphics[height=6cm]{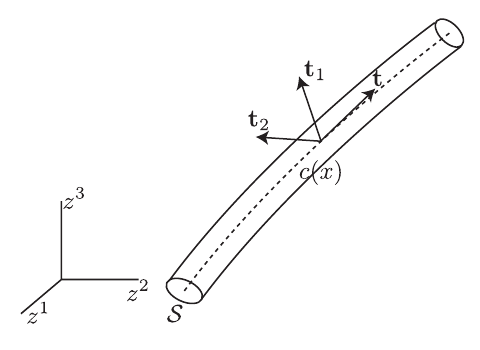}
    \end{center}
\caption{Geometry of the functionally graded rod: (a) reference curve $c(x)$, (b) cross-section $\mathcal{S}$, and (c) the local co-moving orthonormal frame $\{\mathbf{t}_1, \mathbf{t}_2, \mathbf{t}\}$.}
    \label{fig:1}
\end{figure}

We consider a three-dimensional elastic body occupying the domain $\mathcal{V}$, generated by sweeping a planar cross-section $\mathcal{S}$ along a smooth reference curve $c(x)$, where $x$ denotes the arc-length. The cross-section is assumed to be locally orthogonal to the curve, intersecting it at a fixed point, typically chosen as the centroid (see Fig.~\ref{fig:1}). For a functionally graded (FG) rod, the material properties are defined as functions of the transverse coordinates within $\mathcal{S}$.

To facilitate a rigorous asymptotic analysis, we describe the reference configuration using a fixed Cartesian frame $\{ z^1, z^2, z^3\}$ and a co-moving orthonormal triad $\{\mathbf{t}_1, \mathbf{t}_2, \mathbf{t}\}$. Here, $\mathbf{t} = \mathbf{r}'$ is the unit tangent to the central line, and $\mathbf{t}_\alpha$ ($\alpha=1,2$) are unit vectors rigidly associated with the cross-section. The position of any material point is specified by:
\begin{equation}
\label{eq:1}
\mathbf{z}(x^\alpha,x)=\mathbf{r}(x) +\mathbf{t}_\alpha (x)x^\alpha .
\end{equation}
The coordinates $x^\alpha$ span the connected 2D domain $\mathcal{S}$, such that the first moment of area vanishes, $\int_{\mathcal{S}} x^\alpha \dd{a}=0$.

For an anisotropic FG-rod, the dynamic behavior is governed by the action functional expressed in curvilinear coordinates. Following the general formulation for slender rods \cite{berdichevsky2009variational,le1999vibrations}, the action functional reads:
\begin{equation}
\int_{t_0}^{t_1}\int_{\mathcal{V}}\mathcal{L}(x^\alpha ,\dot{\mathbf{w}},\nabla \mathbf{w})\dd{v}\dd{t}=\int_{t_0}^{t_1}\int_0^L\int_{\mathcal{S}}
[T(x^\alpha ,\dot{\mathbf{w}})-W(x^\alpha ,\boldsymbol{\varepsilon })]\sqrt{g} \dd{a}\dd{x}\dd{t},
\label{eq:2}
\end{equation}
where the geometric factor $\kappa = \sqrt{g} = (1 + \omega_\alpha x^\alpha)$ accounts for the rod's initial curvatures $\omega_\alpha$. The kinetic energy density $T$ is defined as:
\begin{equation}
T(x^\alpha,\dot{\mathbf{w}})=\frac{1}{2}\rho (x^\alpha ) (\dot{w}^\alpha
\dot{w}_\alpha +\dot{w}^2),
\label{eq:3}
\end{equation}
where $\rho(x^\alpha)$ represents the spatially varying mass density. The displacement components $w$ and $w_\alpha$ are projected onto the local frame as follows:
\begin{equation}
w=t^i w_i, \quad w_\alpha =t^i_\alpha w_i .
\label{eq:4}
\end{equation}

The strain energy density $W$ for the most general case of anisotropy and material gradation is expressed in terms of the elastic moduli $c^{abcd}(x^\alpha)$ as:
\begin{multline}\label{eq:5}
W(x^\alpha,\boldsymbol{\varepsilon})=\frac{1}{2}c^{abcd}(x^\alpha )\varepsilon _{ab}\varepsilon _{cd}=\frac{1}{2}c^{\alpha \beta \gamma \delta}\varepsilon_{\alpha \beta}\varepsilon_{\gamma \delta}+\frac{1}{2}c^{3333}(\varepsilon_{33})^2+2c^{3 \alpha 3 \beta }\varepsilon_{3 \alpha }\varepsilon_{3 \beta }
\\
+c^{\alpha \beta 33}\varepsilon_{\alpha \beta}\varepsilon_{33}+2c^{\alpha \beta 3 \gamma }\varepsilon_{\alpha \beta}\varepsilon_{3 \gamma }+2c^{3\alpha 33}\varepsilon_{3 \alpha }\varepsilon_{33},
\end{multline}
where the linearized strain tensor $\boldsymbol{\varepsilon}$ is defined by:
\begin{equation}
\label{eq:6}
\boldsymbol{\varepsilon}=\frac{1}{2}(\nabla \mathbf{w}+(\nabla \mathbf{w})^T).
\end{equation}

Hamilton's principle dictates that the actual displacement field $\check{\mathbf{w}}(x^\alpha,x,t)$ is the stationary point of the action functional \eqref{eq:2}. Our objective is to rigorously reduce this 3D functional to an effective 1D model by exploiting the smallness of the parameters $h/R$, $h/l$, and $h/(v\tau)$, where $h$ is the cross-sectional diameter, $R$ is the curvature-twist radius, $l$ is the longitudinal characteristic scale, and $v\tau$ is the characteristic wavelength in dynamics. Condition $h/(v\tau)\ll 1$ means that we consider only low-frequency vibrations of the rod.\footnote{There are extremely interesting high-frequency vibrations of elastic and piezoelectric shells and rods, the study of which was initiated in the pioneering work of Mindlin \cite{mindlin1961high}. The rigorous dimension reduction based on VAM involving also the kinetic energy density has been considered in \cite{le1999vibrations,berdichevsky1980high,le1997high,lee2009dynamic}.}

By introducing the scaled dimensionless coordinates $\zeta^\alpha = x^\alpha/h$, the components of the strain tensor take the following asymptotic forms:
\begin{align}
\varepsilon _{\alpha \beta }&
=\frac{1}{h} w_{(\alpha ,\beta )}, \label{eq:9}
\\
\varepsilon _{3 \alpha }&=\frac{1}{2}[w_{\alpha }^\prime  
-\omega _\alpha w-\varpi  
e^{. \beta }_{\alpha .}w_\beta +\frac{1}{h}(1+h\omega _\beta  
\zeta ^\beta )
w_{,\alpha }+\varpi e^{.\delta }_{\gamma .}  
\zeta ^\gamma w_{\delta ,\alpha }], \label{eq:10}
 \\
\varepsilon _{33}&=(1+h\omega _\beta \zeta ^\beta )
(w^\prime +\omega ^\alpha w_\alpha )
+h\varpi e^{.\delta }_{\gamma .} \zeta ^\gamma  
(w_{\delta }^\prime -\omega _\delta w-\varpi e^{.\kappa  
}_{\delta .}w_\kappa ) , \label{eq:11}
\end{align}
where $\varpi$ denotes the initial twist of the rod. These kinematic relations serve as the foundation for the energetic decomposition and subsequent dimension reduction.

\section{Two-dimensional elastic moduli}
Prior to the application of the variational-asymptotic procedure, it is advantageous to reconfigure the strain energy density into a form that highlights the separation between longitudinal and transverse effects \cite{le1999vibrations}. Given that the derivatives with respect to the transverse coordinates represent the dominant kinematic terms in \eqref{eq:9}--\eqref{eq:11}, we group the energy contributions such that the axial and shear-plane interactions are decoupled. Mathematically, this process corresponds to a Schur complement-like reduction of the 3D stiffness tensor, wherein the degrees of freedom associated with the transverse strains are condensed through energy minimization \cite{cook2007concepts,yu2002on}. Specifically, the total strain energy density \eqref{eq:5} is partitioned into:
\begin{equation}
W=W_\parallel +W_\perp,
\label{eq:12}
\end{equation}
where $W_\parallel (\varepsilon_{33})$ represents the longitudinal energy density, defined by the following minimization over the transverse strain components:
\begin{equation}
W_\parallel (\varepsilon_{33})=\min_{\varepsilon _{\alpha \beta }, \varepsilon _{3\alpha }}W(\varepsilon _{\alpha \beta},\varepsilon_{3\alpha },\varepsilon_{33}).
\label{eq:13}
\end{equation}
The transverse energy density, $W_\perp$, accounts for the remaining portion of the energy associated with the deformation in the cross-sectional plane:
\begin{equation}
\label{eq:14}
W_\perp(\varepsilon _{\alpha \beta},\varepsilon_{3\alpha },\varepsilon_{33})=W(\varepsilon _{\alpha \beta},\varepsilon_{3\alpha },\varepsilon_{33})-W_\parallel (\varepsilon_{33}).
\end{equation}

Direct algebraic reduction yields the explicit forms:
\begin{align}
W_\parallel &=\frac{1}{2}E (\varepsilon _{33})^2, \label{eq:15} \\
W_\perp &=\frac{1}{2}(c^{\alpha \beta \gamma \delta } \gamma _{\alpha \beta }\gamma _{\gamma \delta }+ c^{3\alpha 3\beta }\gamma _\alpha \gamma _\beta + 2c^{\alpha \beta 3\gamma }\gamma _{\alpha \beta } \gamma _\gamma ),
\label{eq:16}
\end{align}
where the reduced strain measures $\gamma_\alpha$ and $\gamma_{\alpha \beta}$ are defined by:
\begin{align}
\gamma_\alpha&=2\varepsilon_{3\alpha }+r_\alpha \varepsilon_{33}, \label{eq:17} \\
\gamma_{\alpha \beta }&=\varepsilon_{\alpha \beta}+r_{\alpha \beta }\varepsilon_{33}. \label{eq:18}
\end{align}

The effective longitudinal Young modulus $E$ and the coupling tensors $r_{\alpha \beta }$, $r_\alpha$ are derived from the 3D anisotropic moduli through the relations:
\begin{gather}
E= \bar{c}^{3333} -r_\alpha \bar{c}^{3\alpha 33}, \quad r_\alpha =s_{\alpha \beta }\bar{c}^{3\beta 33}, \quad \bar{s}_{\alpha \beta}=(\bar{c}^{3\alpha 3\beta })^{-1}, \label{eq:19} \\
\bar{c}^{3333}=c^{3333}-r_{\alpha \beta }c^{\alpha \beta 33}, \quad \bar{c}^{3\alpha 3a}=c^{3\alpha 3a}-r^\alpha _{\mu \nu } c^{\mu \nu 3a}, \label{eq:20} \\
r_{\alpha \beta }=b_{\alpha \beta \mu \nu }c^{\mu \nu 33}, \quad r_{\alpha \beta }^\lambda =b_{\alpha \beta \mu \nu }c^{\mu \nu 3\lambda }, \label{eq:21}
\end{gather}
where $b_{\alpha \beta \mu \nu }$ is the inverse of the transverse stiffness tensor $c^{\alpha \beta \mu \nu }$. It is important to note that the transverse energy $W_\perp$ represents the state where the in-plane and out-of-plane shear stresses vanish. Consequently, $E$ serves as the effective longitudinal stiffness of the anisotropic rod under these relaxed constraints.

The coefficients $c^{\alpha \beta \gamma \delta}$, $c^{3\alpha 3\beta }$, $c^{\alpha \beta 3\gamma }$, $r_{\alpha \beta }$, $r_\alpha $, $r^\lambda_{\alpha \beta }$, and $E$ constitute the independent ``two-dimensional'' (2D) elastic moduli of the theory. These moduli inherit the following symmetry properties from the original 3D tensor:
\begin{gather}
c^{\alpha \beta \gamma \delta }= c^{\beta \alpha \gamma \delta }= c^{\alpha \beta \delta \gamma }= c^{\gamma \delta \alpha \beta }, \label{eq:22} \\
c^{3\alpha 3\beta } = c^{3\beta 3\alpha } ,\quad c^{\alpha \beta 3\gamma }=c^{\beta \alpha 3\gamma }=c^{3\gamma \alpha \beta}, \label{eq:23} \\
r_{\alpha \beta }=r_{\beta \alpha },\quad r^\lambda_{\alpha \beta }=r^\lambda_{\beta \alpha }. \label{eq:24}
\end{gather}

In general, these 2D moduli vary across the section $\mathcal{S}$ due to both physical material gradation and the geometric shifters associated with the curvilinear frame. However, for a functionally graded rod with moderate curvature, the moduli can be approximated by their physical values in a local Cartesian frame:
\begin{equation}
A(\zeta ^\alpha)=A_0(\zeta ^\alpha)+O(h/R) A_0(\zeta ^\alpha),
\label{eq:26}
\end{equation}
where $A_0(\zeta ^\alpha)$ represents the primary material inhomogeneity. This study specifically considers two symmetry cases where the number of independent constants is reduced: (i) symmetry with respect to the plane perpendicular to the central line, and (ii) transversal isotropy.

\section{Asymptotic analysis of the action functional}

The construction of a one-dimensional theory for functionally graded rods relies on the systematic asymptotic analysis of the 3D action functional. Given the slender geometry of the rod, the first and second steps in this asymptotic analysis provide the rod kinematics ``in average'', while the subsequent step introduces the local cross-sectional warping as a perturbation of the average rod motion.

To bypass the recursive steps of finding leading-order kinematics and subsequent asymptotic corrections—standard in Berdichevsky’s original iterative procedure—we represent the displacement field $\mathbf{w}$ using a structured Ansatz that encapsulates the average motion and local warping corrections directly:
\begin{align}
w_\alpha &=u_\alpha (x,t)-he_{\alpha \beta }\phi  
(x,t)\zeta ^\beta +hy _\alpha (\zeta ^\alpha ,x,t), \label{eq:29}
 \\
w&=u(x,t)-he^{.\beta }_{\alpha .}\phi _\beta (x,t)
\zeta ^\alpha +hy(\zeta ^\alpha ,x,t),
\label{eq:30} 
\end{align}
where the functions $u_\alpha$ and $u$ denote the translational displacements of the cross-section, and $\phi$ represents the average twist angle. The auxiliary functions $\phi_\alpha$ are determined by the slope of the central line and the initial geometry:
\begin{equation}
\phi _\alpha =-e_{\alpha .}^{.\beta }u^\prime _\beta
-\varpi u_\alpha +e_{\alpha .}^{.\beta }\omega _\beta u.
\label{eq:31}
\end{equation} 
To ensure the decomposition is unique, the warping functions $y_\alpha$ and $y$ are required to satisfy the following normalization conditions:
\begin{gather}
\langle y_\alpha \rangle =0,\quad \langle y \rangle =0,
\quad e^{\alpha \beta } \langle y_{\alpha ,\beta }\rangle =0,
\label{eq:32} 
\end{gather}
where $\langle .\rangle $ denotes the integration over 2D domain $\bar{\mathcal{S}}$ defined in terms of the dimensionless cross-sectional coordinates $\zeta^\alpha$. 

If the warping functions vanish, the deformation state of the 1D model is characterized by the elongation $\gamma$, the bending curvatures $\Omega_\alpha$, and the twist $\Omega$, defined as:
\begin{align}
\gamma &=u^\prime +\omega ^\alpha u_\alpha , \label{eq:33}
 \\
\Omega _\alpha &=-e_{\alpha .}^{. \beta }\phi ^\prime _\beta  
-\varpi \phi _\alpha +e_{\alpha .}^{. \beta } \omega _\beta  
\phi ,
\label{eq:34} \\
\Omega &=\phi ^\prime +\omega ^\alpha \phi _\alpha . \label{eq:35}
\end{align}

Substituting the displacement field into the kinematic relations, we find that the strain components are dominated by the transverse gradients of the warping functions and the global 1D strain measures:
\begin{equation}
\varepsilon _{\alpha \beta }=y_{(\alpha ,\beta )},\quad 
2\varepsilon _{3\alpha }=
y_{,\alpha }-he_{\alpha \beta }\Omega \zeta ^\beta ,
\quad \varepsilon _{33}=\gamma + h\Omega_\sigma \zeta ^\sigma . \label{eq:40}
\end{equation}

The determination of the optimal warping functions reduces to solving a two-dimensional variational problem on the cross-section. We define the average transverse energy density, $\Phi_\perp$, as the minimum of the transverse energy functional:
\begin{equation}
\Phi _\perp (\gamma,\Omega,\Omega_\alpha)=h^2 \min_{y_\alpha ,y\in \eqref{eq:32}} \langle W_\perp (\gamma _{\alpha \beta },
\gamma _\alpha )\rangle ,
\label{eq:41} 
\end{equation}
where the reduced strain measures $\gamma_{\alpha \beta}$ and $\gamma_\alpha$ incorporate the 1D kinematic variables:
\begin{align}
\gamma _{\alpha \beta }&=y_{(\alpha ,\beta )}+
r_{\alpha \beta }(\gamma +h\Omega _\sigma \zeta ^\sigma ),
\label{eq:42} \\
\gamma _\alpha &=y_{,\alpha }-he_{\alpha \beta }\Omega \zeta 
^\beta +r_{\alpha }(\gamma +h\Omega _\sigma \zeta ^\sigma ).
\label{eq:43} 
\end{align}

Upon solving the cross-sectional problem \eqref{eq:41}, the total 1D strain energy density $\Phi$ is obtained by summing the longitudinal and transverse contributions:
\begin{equation}
\label{eq:44}
\Phi(\gamma,\Omega,\Omega_\alpha)= \Phi _\parallel (\gamma,\Omega_\alpha) + \Phi _\perp (\gamma,\Omega,\Omega_\alpha).
\end{equation}
The longitudinal part $\Phi_\parallel$ is a quadratic form determined by the effective axial modulus $E$ and the cross-sectional geometry:
\begin{equation}
\Phi _\parallel =\frac{h^2}{2}\langle E(\gamma +
h\Omega_\alpha \zeta ^\alpha )^2\rangle 
=\frac{1}{2}(E_\parallel \gamma ^2+2B_{\alpha \parallel} \gamma \Omega^\alpha +B_{\alpha \beta \parallel }\Omega^\alpha \Omega ^\beta ).
\label{eq:46}
\end{equation}

Finally, the 1D kinetic energy $\Theta$ is derived by integrating the 3D kinetic energy density over the section, accounting for the mass distribution of the functionally graded material:
\begin{equation}
\Theta =\frac{h^2}{2}[ \langle \rho \rangle (\dot{u}^2
+\dot{u}^\alpha \dot{u}_\alpha )-2h\langle \rho e_{\alpha \beta} \zeta^\beta \rangle \dot{u}_\beta \dot{\phi }+h^2\langle \rho \delta_{\alpha \beta} \zeta ^\alpha \zeta^\beta \rangle \dot{\phi }^2] .
\label{eq:47}
\end{equation}
This formulation establishes the governing 1D energy functional, where the local material gradation is rigorously homogenized into effective rod stiffnesses and inertia coefficients.

\section{Dual cross-sectional problem}

For functionally graded rods exhibiting general anisotropy, the cross-sectional problem established in \eqref{eq:41} typically necessitates a numerical solution. To ensure the mathematical reliability of these results and to facilitate precise error control, we develop the dual cross-sectional problem. This formulation provides a complementary variational principle, enabling the determination of both upper and lower bounds for the effective one-dimensional stiffness coefficients.\footnote{The first idea in this direction has been proposed for the cross-sectional problems of functionally graded elastic isotropic beam \cite{le2025asymptotically}.} Since the 2D metric tensor in the cross-section plane is the Kronecker delta, $\delta_{\alpha \beta}$, raising or lowering Greek indices do not affect the components of vectors and tensors at all. For this reason, we shall write all equations in Sections 5 and 6 in covariant form with vectors and tensors having lower indices. The summation convention of summing up over repeated indices within their range applies also in this case.

We initiate the dualization by applying the Young-Fenchel transformation to the transverse energy density $W_\perp$, which yields the complementary energy density $W^*_\perp (\tau_{\alpha \beta},\tau_\alpha)$:
\begin{equation}
\label{eq:48}
W^*_\perp (\tau_{\alpha \beta},\tau_\alpha)=\max_{\gamma_{\alpha \beta},\gamma_\alpha}[\tau_{\alpha \beta} \gamma_{\alpha \beta}+\tau_\alpha \gamma_\alpha -W_\perp (\gamma_{\alpha \beta},\gamma_\alpha)].
\end{equation}
The resulting complementary density is a positive definite quadratic form of the transverse stresses:
\begin{equation}
W^*_\perp (\tau_{\alpha \beta},\tau_\alpha) =\frac{1}{2}(\bar{s}_{\alpha \beta \gamma \delta } \tau_{\alpha \beta }\tau_{\gamma \delta }+
\bar{s}_{\alpha \beta } \tau_\alpha \tau_\beta +
2\bar{s}_{\alpha \beta \gamma} \tau_{\alpha \beta }\tau_\gamma ),
\label{eq:49}
\end{equation}
where the 2D elastic compliances $\bar{s}_{\alpha \beta \gamma \delta }$ and $\bar{s}_{\alpha \beta \gamma}$ are:
\begin{equation}
\bar{s}_{\alpha \beta \gamma \delta }=b_{\alpha \beta \gamma \delta }+\bar{s}_{\lambda \mu } r_{\lambda \alpha \beta }r_{\mu \gamma \delta },\quad
\bar{s}_{\alpha \beta \gamma}=-\bar{s}_{\gamma \mu} r_{\mu \alpha \beta}, \label{eq:50}
\end{equation}
with $\bar{s}_{\alpha \beta}$, $b_{\alpha \beta \gamma \delta}$ and $r_{\lambda \alpha \beta }$ being defined in \eqref{eq:20}. 

By exploiting the strong duality of the variational problem, the minimization in \eqref{eq:41} can be transformed into a maximization over a statically admissible stress field. This allows us to express the average transverse energy density as:
\begin{multline}
\label{eq:59}
\Phi _\perp =h^2 \max_{\tau_{\alpha \beta}, \tau_\alpha} \langle \tau_{\alpha \beta}r_{\alpha \beta}(\gamma +h\Omega_\sigma \zeta_\sigma )-\tau_\alpha e_{\alpha \beta}h\Omega \zeta_\beta  
\\
+\tau_\alpha r_\alpha (\gamma +h\Omega_\sigma \zeta_\sigma )-W^*_\perp (\tau_{\alpha \beta }, \tau_\alpha ) \rangle,
\end{multline}
where the transverse stress components must satisfy the local equilibrium and boundary conditions:
\begin{eqnarray}
&\tau_{\alpha \beta, \beta} = 0,\quad \tau_{\alpha \beta}n_\beta=0 \text{ at $\partial \bar{\mathcal{S}}$}, \label{eq:57} \\
&\tau_{\alpha , \alpha} = 0,\quad \tau_{\alpha }n_\alpha=0 \text{ at $\partial \bar{\mathcal{S}}$} . \label{eq:58}
\end{eqnarray}

To satisfy these constraints identically, we introduce a vector stress function $\psi_\alpha$ and a scalar stress function $\psi$, such that the stresses are represented as:
\begin{equation}
\label{eq:60}
\tau_{\alpha \beta}=e_{\alpha \gamma }e_{\beta \delta}\psi_{\gamma ,\delta }, \quad \tau_\alpha =e_{\alpha \beta} \psi_{,\beta }.
\end{equation}
The functions $\psi_\alpha$ and $\psi$ are required to vanish on the boundary $\partial \bar{\mathcal{S}}$ to meet the traction-free boundary conditions:
\begin{equation}
\label{eq:61}
\psi_\alpha =0, \quad \psi=0 \quad \text{ at $\partial \bar{\mathcal{S}}$}.
\end{equation}
The symmetry of the stress tensor further imposes the constraint $e_{\alpha \beta}\psi_{\alpha ,\beta}=0$. Enforcing this via a penalty approach leads to the unconstrained maximization problem:
\begin{multline}
\label{eq:63}
\Phi _\perp =h^2 \max_{\psi_{\alpha }, \psi \in \eqref{eq:61}} \langle e_{\alpha \gamma }e_{\beta \delta}\psi_{\gamma ,\delta }r_{\alpha \beta}(\gamma +h\Omega_\sigma \zeta_\sigma )-\psi_{,\alpha }h\Omega \zeta_\alpha
\\
+e_{\alpha \beta} \psi_{,\beta } r_\alpha (\gamma +h\Omega_\sigma \zeta_\sigma )-W^*_\perp (e_{\alpha \gamma }e_{\beta \delta}\psi_{\gamma ,\delta },  
e_{\alpha \beta} \psi_{,\beta }) -\frac{1}{2}\theta (e_{\alpha \beta}\psi_{\alpha ,\beta})^2 \rangle ,
\end{multline}
where $\theta$ serves as the penalty parameter. This dual formulation acts as a critical validation mechanism; the convergence of the results from \eqref{eq:41} and \eqref{eq:63} ensures that the effective one-dimensional stiffnesses are accurately captured, effectively bracketing the true transverse energy even for rods with high-contrast material gradation.

\section{Numerical solution of cross-sectional problems}

\subsection{Finite element implementation of the direct and dual problems}

The determination of the warping functions and stress functions requires the solution of the boundary value problems derived in Sections 4 and 5. To accommodate arbitrary cross-sectional geometries and complex material gradations, we cast these systems into a finite element framework using the MATLAB PDE Toolbox. The governing equations for both the direct and dual formulations can be generalized into a system of elliptic PDEs in divergence form:
\begin{align}
&-\nabla \cdot (\mathbf{c} \otimes \nabla \mathbf{u}) + \mathbf{a}\mathbf{u} = \mathbf{f} \quad \text{in } \bar{\mathcal{S}}, \label{eq:74} \\
&\mathbf{n} \cdot (\mathbf{c} \otimes \nabla \mathbf{u}) = \mathbf{g} \quad \text{on } \partial \bar{\mathcal{S}}, \label{eq:75}
\end{align} 
where $\mathbf{u}$ represents the vector of unknowns. This system is characterized by the matrices $\mathbf{c}$ and the vectors $\mathbf{f}, \mathbf{g}$, which account for the coupled effects of anisotropy and material gradation. The explicit, expanded forms of these matrices and vectors, which are essential for numerical implementation but algebraically intensive, are provided in Appendix A.

The boundary-value problems \eqref{eq:74}-\eqref{eq:75} are solved by providing the specific $\mathbf{c}$-matrices, source vectors ($\mathbf{f}, \mathbf{g}$) (see Appendix A), and the cross-sectional domain geometry to the MATLAB PDE Toolbox \citep{mathworks1995partial}. Upon obtaining the discretized field solutions, the energy functionals are evaluated using the high-order 'integral2' numerical integration scheme. A defining feature of this dual approach is the built-in error quantification: because the primal and dual problems yield rigorous upper and lower bounds for the transverse energy $\Phi_\perp$, respectively, the exactness of the numerical result is guaranteed by the convergence of these two values. The computational process, including adaptive mesh refinement, continues until the gap between the energy bounds is within a specified tolerance.

Exploiting the linearity of the underlying equations, the warping and stress functions are linear combinations of the 1D strain measures $\gamma, \Omega$, and $\Omega_\alpha$. Consequently, the minimized transverse energy takes the form of a general quadratic functional:
\begin{equation}
\label{eq:96}
\Phi_\perp=\frac{1}{2}\left[E_\perp \gamma ^2+B_{\alpha \beta \perp} \Omega_\alpha \Omega_\beta+C \Omega^2+2\gamma (B_{\alpha \perp} \Omega_\alpha +D \Omega )+2D_\alpha \Omega_\alpha \Omega\right] . 
\end{equation}

A significant computational advantage of the dual framework is its ability to determine effective 1D stiffnesses without requiring the derivatives of the numerical solution. This property relaxes the regularity requirements on the shape functions, necessitating only $C^0$-continuity for the stress functions. By invoking Clapeyron’s identity, the average transverse energy density is reconstructed directly from the stress-function maximizers $(\check{\psi}_\alpha, \check{\psi})$:
\begin{equation}
\Phi_\perp = \frac{1}{2}h^2 \langle -e_{\alpha \gamma }e_{\beta \delta}\check{\psi}_\gamma (r_{\alpha \beta}(\gamma +h\Omega_\sigma \zeta_\sigma ))_{,\delta }+2\check{\psi}h\Omega 
-e_{\alpha \beta} \check{\psi} (r_\alpha (\gamma +h\Omega_\sigma \zeta_\sigma ))_{,\beta }\rangle. \label{eq:97}
\end{equation}

To extract the individual components of the $4 \times 4$ 1D stiffness matrix, we employ a systematic perturbation method. Diagonal coefficients are isolated by setting the corresponding 1D strain measure to unity while nullifying all others. For the off-diagonal (coupling) terms, we utilize the stress functions calculated from a unit load in one strain measure to evaluate the energy density under a second unit strain measure. This robust extraction procedure, established for isotropic beams in \cite{le2025asymptotically}, is here generalized for the anisotropic case.

The numerical framework developed in this subsection is sufficiently robust to accommodate any combination of cross-sectional geometry, material anisotropy, and spatial gradation. In the subsequent subsections, we demonstrate the utility of this approach by analyzing FG-rods with two specific material symmetries: (i) hexagonal symmetry (transversal isotropy) and (ii) rhombohedral symmetry, both with their primary axes aligned with the rod's central tangent.

\subsection{Numerical results: transversal isotropy}

For functionally graded materials possessing transversal isotropy, the symmetry dictates that all 2D tensors with an odd number of Greek indices--specifically $c_{\alpha \beta 3 \gamma}$, $r_\alpha$, and $r_{\alpha \beta \lambda}$--vanish. Moreover, the fourth-rank tensor $c_{\alpha \beta \gamma \delta}$ is represented by:
\begin{equation}
c_{\alpha \beta \gamma \delta}=\lambda \delta_{\alpha \beta}\delta_{\gamma \delta}+\mu (\delta_{\alpha \gamma}\delta_{\beta \delta}+\delta_{\alpha \delta}\delta_{\beta \gamma}),
\end{equation}
while the second-rank tensors take a spherical form:
\begin{equation}
\label{eq:98}
\bar{c}_{3\alpha 3\beta}=G\delta_{\alpha \beta},\quad r_{\alpha \beta}=\nu \delta_{\alpha \beta}.
\end{equation}
In this case, the material behavior is governed by five independent elastic constants: the longitudinal Young's modulus $E$ (or $E_3$), the out-of-plane shear modulus $G$ ($G_{13}$), the out-of-plane Poisson's ratio $\nu$ ($\nu_{31}$), and the in-plane Lamé constants $\lambda$ and $\mu$. Unlike purely isotropic media, $\lambda$ and $\mu$ are independent of $E$ and $\nu$. We characterize the in-plane properties using the in-plane Young's modulus $E_1$ and Poisson's ratio $\nu_1$ via:
\begin{equation}
\label{eq:99}
\lambda=\frac{E_1\nu_1}{1-\nu_1^2},\quad \mu=\frac{E_1}{2(1+\nu_1)}.
\end{equation}
Thermodynamic stability (positive-definiteness of the stored energy density) requires:
\begin{equation}
\label{eq:100}
E>0,\quad G>0,\quad E_1>0,\quad -1<\nu_1<1,\quad \nu^2<\frac{E_1}{2E}(1-\nu_1^2).
\end{equation}
The last inequality is particularly significant as it constrains the out-of-plane Poisson's ratio based on the anisotropy ratio $E_1/E$.

The vanishing of $r_{\alpha \beta \lambda}$ and $r_\alpha$ facilitates the decoupling of the cross-sectional analysis into two independent problems: (i) an anti-plane shear problem for the warping $y$ (dual to $\psi$), and (ii) a plane strain problem for the warping vector $\mathbf{u}=(y_1,y_2)^T$ (dual to $\psi_1,\psi_2$). Despite this decoupling, the presence of five moduli introduces a significantly broader spectrum of material responses than those found in isotropic cases.

The anti-plane problem is resolved by extremizing the following dual functionals:
\begin{equation}
\label{eq:101}
\begin{split}
&h^2 \min_{\langle y\rangle =0} \Bigl\langle \frac{1}{2}G(\boldsymbol{\zeta}) (y_{,\alpha }-e_{\alpha \beta}h\Omega \zeta_\beta)(y_{,\alpha }-e_{\alpha \gamma}h\Omega \zeta_\gamma)\Bigr\rangle \\
& =h^2\max_{\psi |_{\partial \bar{\mathcal{S}}}=0} \Bigl\langle -\psi _{,\alpha }h\Omega \zeta_\alpha -\frac{1}{2G(\boldsymbol{\zeta})} \psi_{,\alpha } \psi_{,\alpha } \Bigr\rangle .
\end{split}
\end{equation} 
Given the linear dependence of the minimizer/maximizer on $h\Omega$, the torsional energy is defined as $\frac{1}{2}g h^4\Omega^2$. The normalized torsional stiffness $g$ is extracted from the unit-load dual problem:
\begin{equation}
\label{eq:102}
g=\min_{\langle y\rangle =0} \langle G(\boldsymbol{\zeta}) (y_{,\alpha }-e_{\alpha \beta} \zeta_\beta)(y_{,\alpha }-e_{\alpha \gamma}\zeta_\gamma)\rangle =\max_{\psi |_{\partial \bar{\mathcal{S}}}=0} \langle -2\psi _{,\alpha }\zeta_\alpha -\frac{1}{G(\boldsymbol{\zeta})} \psi_{,\alpha } \psi_{,\alpha } \rangle .
\end{equation}

For our numerical simulations, we consider a rectangular cross-section with aspect ratio $a$ (width/height). The material properties are assumed to vary along the $\zeta_2$ axis following a power-law distribution \citep{reddy1999axisymmetric}:
\begin{align}
&\{E, E_1, G, \nu, \nu_1\}(\zeta_2)=P_L (1/2-\zeta_2)^\delta + P_U (1-(1/2-\zeta_2)^\delta ),
\label{eq:103-107}
\end{align}
where $P_L$ and $P_U$ denote properties at the lower ($\zeta_2 = -1/2$) and upper ($\zeta_2 = 1/2$) surfaces, respectively. This model effectively represents a two-phase FG composite where both phases (e.g., matrix and fibers) are transversally isotropic. We normalize the moduli by the upper-side longitudinal modulus $E_U$:
\begin{align}
\hat{E}(\zeta_2) &= \epsilon (1/2-\zeta_2)^\delta + (1-(1/2-\zeta_2)^\delta ), \quad \epsilon = E_L/E_U \label{eq:108} \\
\hat{G}(\zeta_2) &= \gamma_L (1/2-\zeta_2)^\delta + \gamma_U(1-(1/2-\zeta_2)^\delta ), \notag
\\ 
&\gamma_L = G_L/E_U,\quad \gamma_U = G_U/E_U \label{eq:109} \\
\hat{E}_1(\zeta_2) &= \epsilon_{1L} (1/2-\zeta_2)^\delta + \epsilon_{1U}(1-(1/2-\zeta_2)^\delta ), \notag
\\
&\epsilon_{1L} = E_{1L}/E_U,\quad \epsilon_{1U} = E_{1U}/E_U \label{eq:110}
\end{align}
The Poisson's ratios $\nu$ and $\nu_1$ remain in their original dimensionless form. The resulting torsional stiffness is expressed as $g=E_U \hat{g}$, where $\hat{g}$ is determined by the normalized shear modulus $\hat{G}(\zeta_2)$ and the section aspect ratio $a$:
\begin{equation}
\label{eq:112}
\hat{g} =\max_{\psi |_{\partial \bar{\mathcal{S}}}=0} \langle -2\psi _{,\alpha }\zeta_\alpha -\frac{1}{\hat{G}(\boldsymbol{\zeta})} \psi_{,\alpha } \psi_{,\alpha } \rangle =2\langle \check{\psi}\rangle .
\end{equation}

\begin{figure}[htb!]
    \centering
    \includegraphics[width=0.45\textwidth]{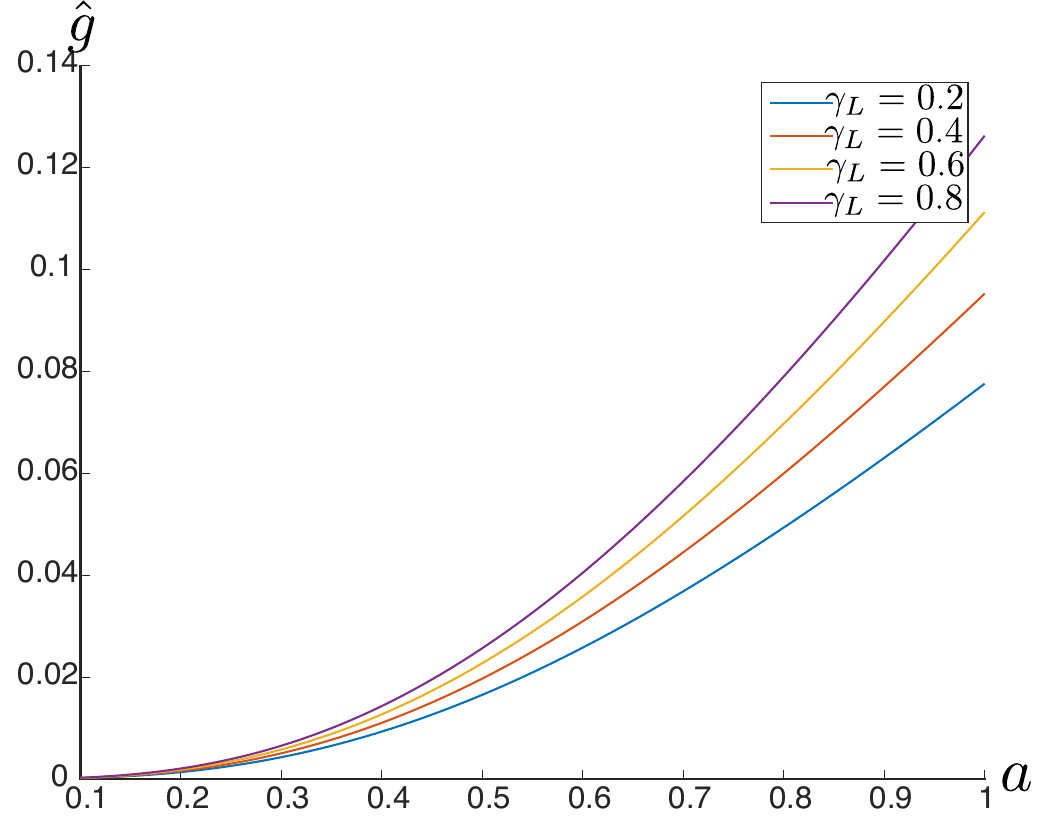}
    \includegraphics[width=0.45\textwidth]{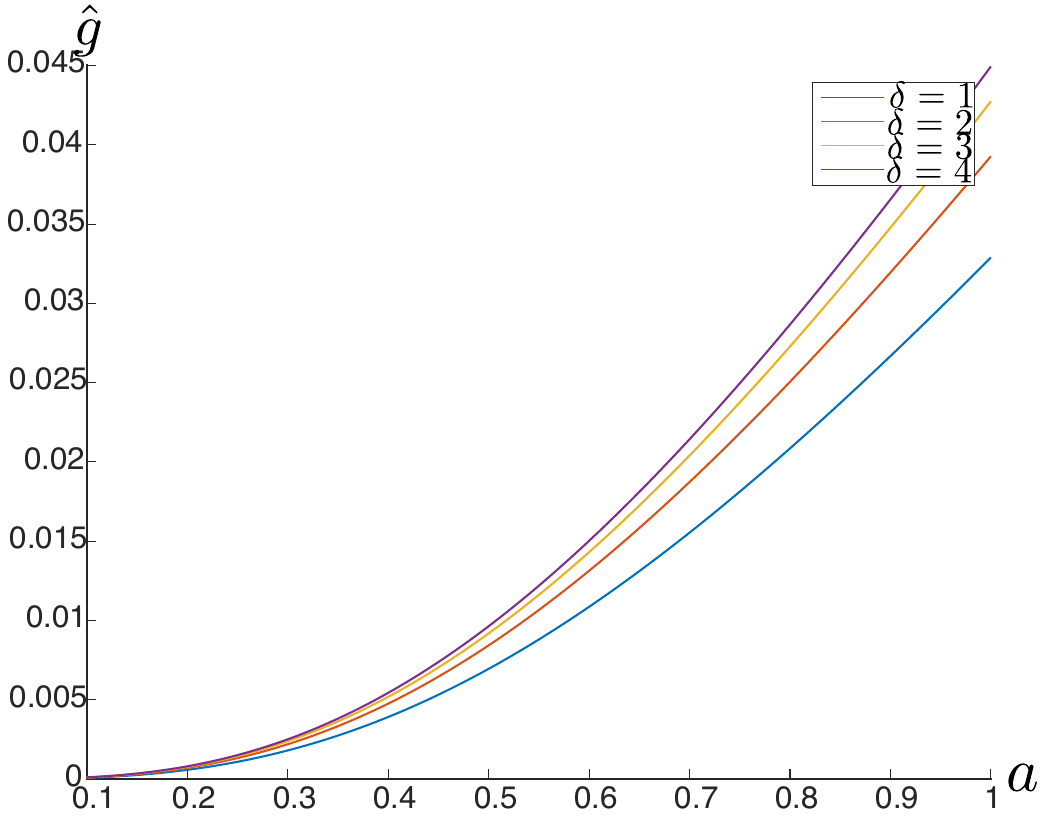}
    \caption{Normalized torsional stiffness $\hat{g}$ as a function of the aspect ratio $a$: (i) at fixed $\delta=1$, $\gamma_U=1$ and four different $\gamma_L=0.2,0.4,0.6,0.8$ (left), and (ii) at fixed $\gamma_L=0.1$, $\gamma_U=0.4$ and four different $\delta=1,2,3,4$ (right).}
    \label{fig:Tstiffness}
\end{figure}

As shown in Fig.~\ref{fig:Tstiffness}, $\hat{g}$ increases monotonically with the aspect ratio $a$, confirming that wider profiles enhance torsional resistance. Moreover, higher values of $\gamma_L$ (stiffer base phase) and larger gradient indices $\delta$ (steeper gradation) contribute positively to the overall torsional stiffness.

\begin{figure}[htb!]
    \centering
    \includegraphics[width=0.6\textwidth]{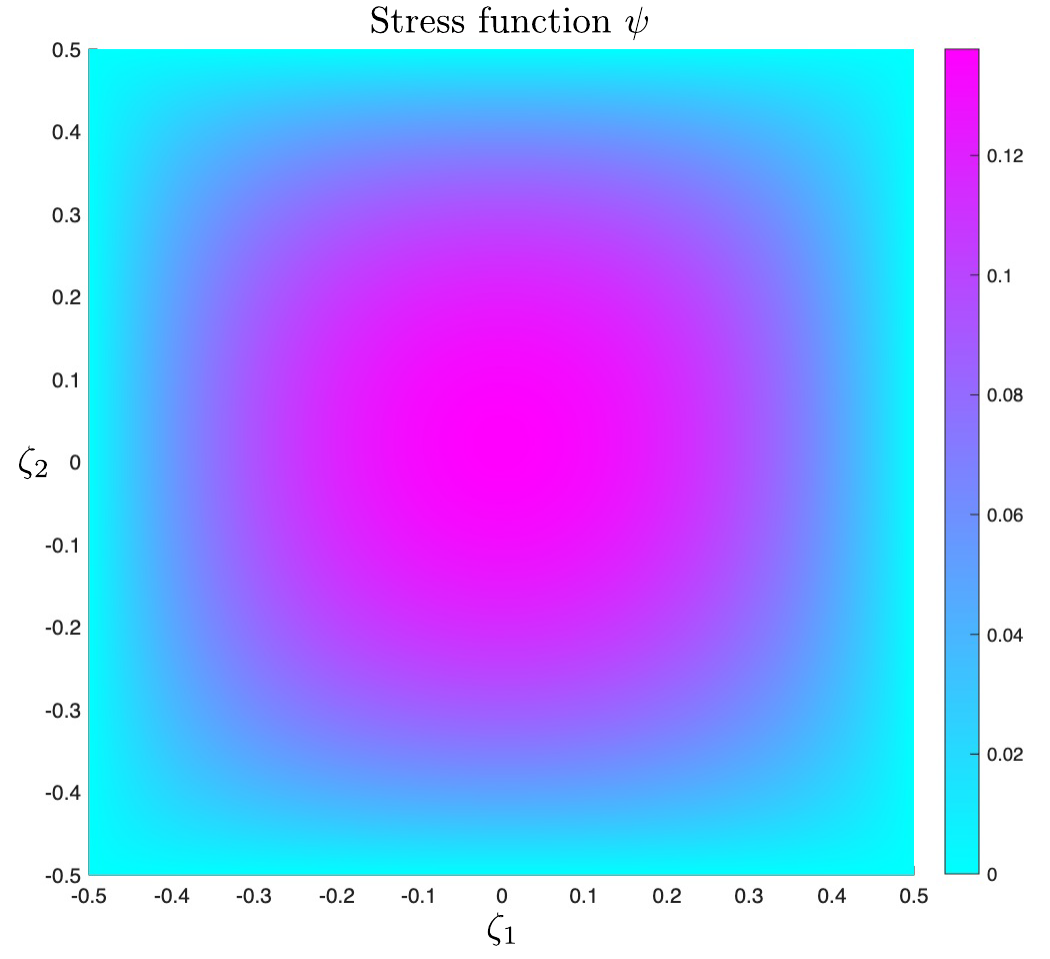}
    \caption{Numerical solution of the stress function $\check{\psi}$ for the anti-plane problem ($a=1$, $\delta=2$, $\gamma_U=1$, $\gamma_L=0.8$). The distribution characterizes the torsional shear stress state across the functionally graded cross-section.}
    \label{fig:StressF1}
\end{figure}

Fig.~\ref{fig:StressF1} displays the contour distribution of the stress function $\check{\psi}$ obtained by solving the dual problem \eqref{eq:102} for the specified material parameters. This plot visually captures the localization of shear resistance within the graded section. Following the dual variational principle, the normalized torsional stiffness $\hat{g}$ is computed as $2\langle \check{\psi} \rangle$, representing twice the average value of the stress function over the dimensionless domain $\bar{\mathcal{S}}$.

For the plane strain component, the normalized $\mathbf{c}$-matrix and driving vectors $\mathbf{f}, \mathbf{g}$ are assembled using the modulated in-plane properties:
\begin{equation}
\label{eq:113}
\mathbf{c}_{p} = \begin{pmatrix}
 \frac{\hat{E}_1}{1-\nu_1^2}  & 0 & 0 & \frac{\hat{E}_1\nu_1}{1-\nu_1^2} \\
 0 & \frac{\hat{E}_1}{2(1+\nu_1)} & \frac{\hat{E}_1}{2(1+\nu_1)} & 0 \\
 0 & \frac{\hat{E}_1}{2(1+\nu_1)} & \frac{\hat{E}_1}{2(1+\nu_1)} & 0 \\
 \frac{\hat{E}_1\nu_1}{1-\nu_1^2} & 0 & 0 & \frac{\hat{E}_1}{1-\nu_1^2}
\end{pmatrix}, \,
\mathbf{f}_{p} = \begin{pmatrix}
   \left(\frac{\hat{E}_1\nu}{1-\nu_1}(\gamma+h\Omega_\beta y_\beta)\right)_{,1}   \\
   \left(\frac{\hat{E}_1\nu}{1-\nu_1}(\gamma+h\Omega_\beta y_\beta)\right)_{,2}  
\end{pmatrix}.
\end{equation}
Crucially, $c_{1133} = c_{2233} = \frac{\hat{E}_1\nu}{1-\nu_1} \neq \lambda$, reducing to the isotropic case only when specific relations between moduli are met. The corresponding dual plane strain matrix is:
\begin{equation}
\label{eq:118}
\mathbf{c}^*_{p} = \begin{pmatrix}
 \frac{1}{\hat{E}_1}  & 0 & 0 & -\frac{\nu_1}{\hat{E}_1} \\
 0 & \frac{1+\nu_1}{2\hat{E}_1}+\theta & \frac{1+\nu_1}{2\hat{E}_1}-\theta & 0 \\
 0 & \frac{1+\nu_1}{2\hat{E}_1}-\theta & \frac{1+\nu_1}{2\hat{E}_1}+\theta & 0 \\
 -\frac{\nu_1}{\hat{E}_1} & 0 & 0 & \frac{1}{\hat{E}_1}
\end{pmatrix}.
\end{equation}

By extracting the normalized stiffness coefficients $\hat{e}_\perp=E_\perp/(E_Uh^2)$, $\hat{b}_{\alpha \perp }=B_{\alpha \perp }/(E_Uh^3)$, and $\hat{b}_{\alpha \beta \perp}=B_{\alpha \beta \perp }/(E_Uh^4)$ via the Clapeyron identity \eqref{eq:97}, we evaluate the total 1D normalized stiffnesses. Our results indicate that while several transverse corrections are negligible, $\hat{b}_{2\perp}$ and $\hat{b}_{22\perp}$ must be retained. To quantify these effects, we define the percentage contributions $\eta_2$ and $\eta_{22}$ as:
\begin{equation}
\label{eq:131}
\eta_2=\frac{\hat{b}_{2\perp} }{\hat{b}_{2\parallel} + \hat{b}_{2\perp }}, \quad \eta_{22}=\frac{\hat{b}_{22\perp} }{\hat{b}_{22\parallel} + \hat{b}_{22\perp }},
\end{equation}
where the longitudinal components are normalized as $\hat{b}_{\alpha \parallel}=B_{\alpha \parallel}/(E_Uh^3)$ and $\hat{b}_{\alpha \beta \parallel}=B_{\alpha \beta \parallel }/(E_Uh^4)$.

\begin{figure}[!htb]
\centering
\includegraphics[width=0.49\textwidth]{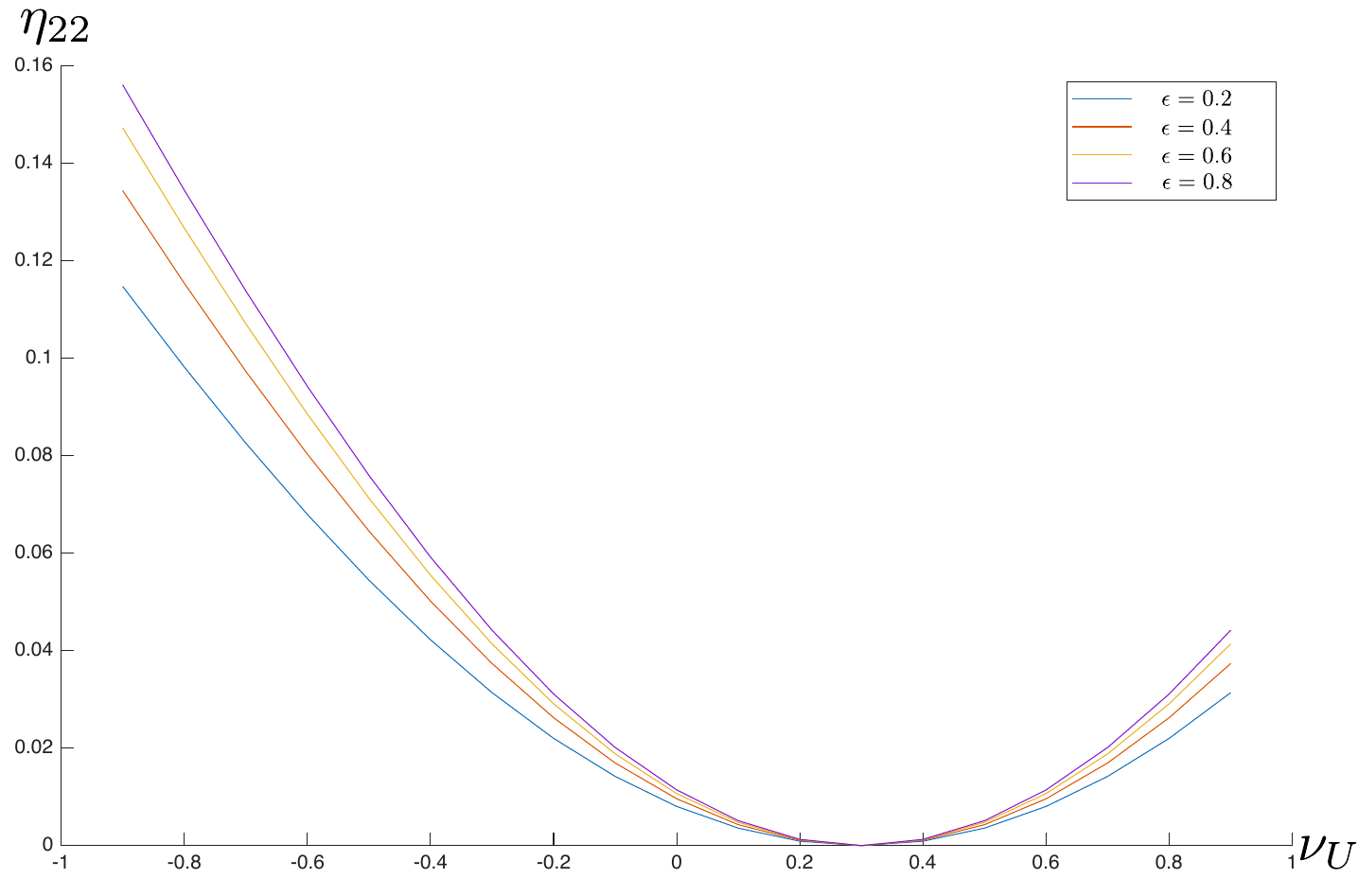}
\includegraphics[width=0.49\textwidth]{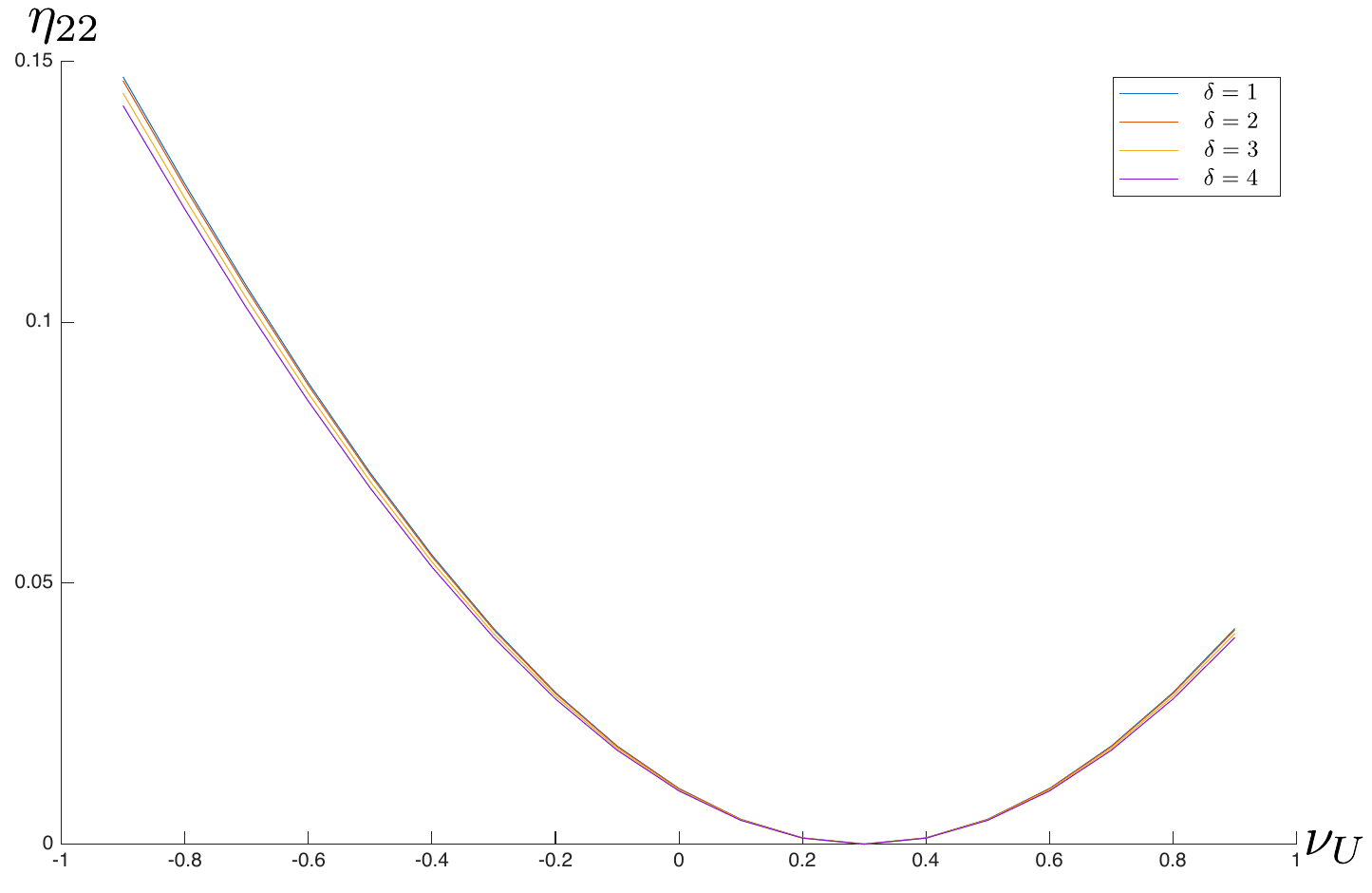} 
\caption{Percentage contribution of the transverse bending stiffness $\eta_{22}$ as a function of the upper-side Poisson's ratio $\nu_U$: (left) for fixed $a=10$, $\delta=4$, $\nu_L=0.3$, and varying $\epsilon=0.2,0.4,0.6,0.8$; (right) for fixed $a=10$, $\epsilon =0.5$, $\nu_L=0.3$, and varying $\delta=1,2,3,4$. The remaining parameters are: $\epsilon_{1L}=2\epsilon$, $\epsilon_{1U}=2$, $\nu_{1L}=0.3$, $\nu_{1U}=0.4$.}
\label{fig:percentage1}
\end{figure}

\begin{figure}[!htb]
\centering
\includegraphics[width=0.49\textwidth]{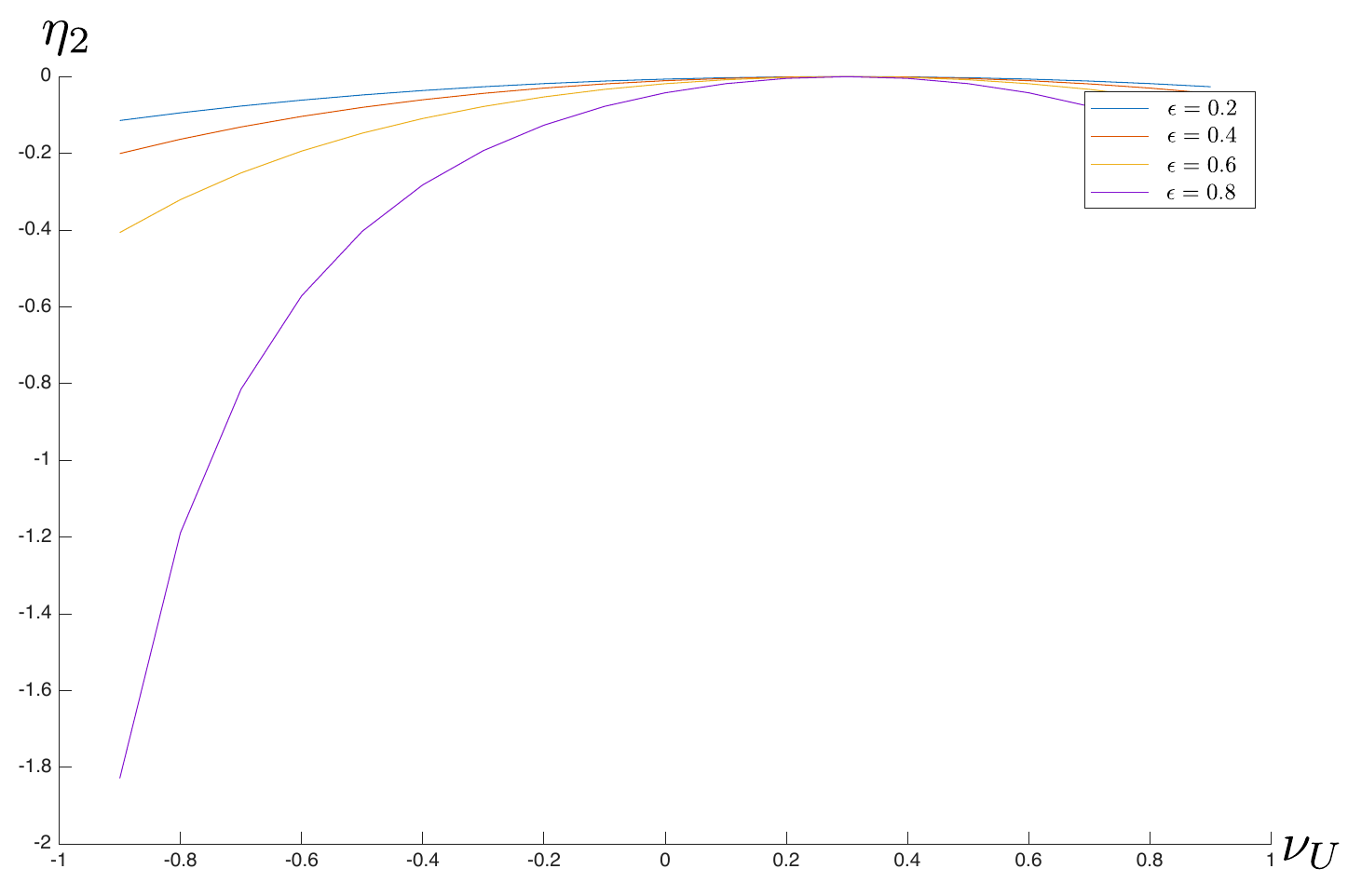}
\includegraphics[width=0.49\textwidth]{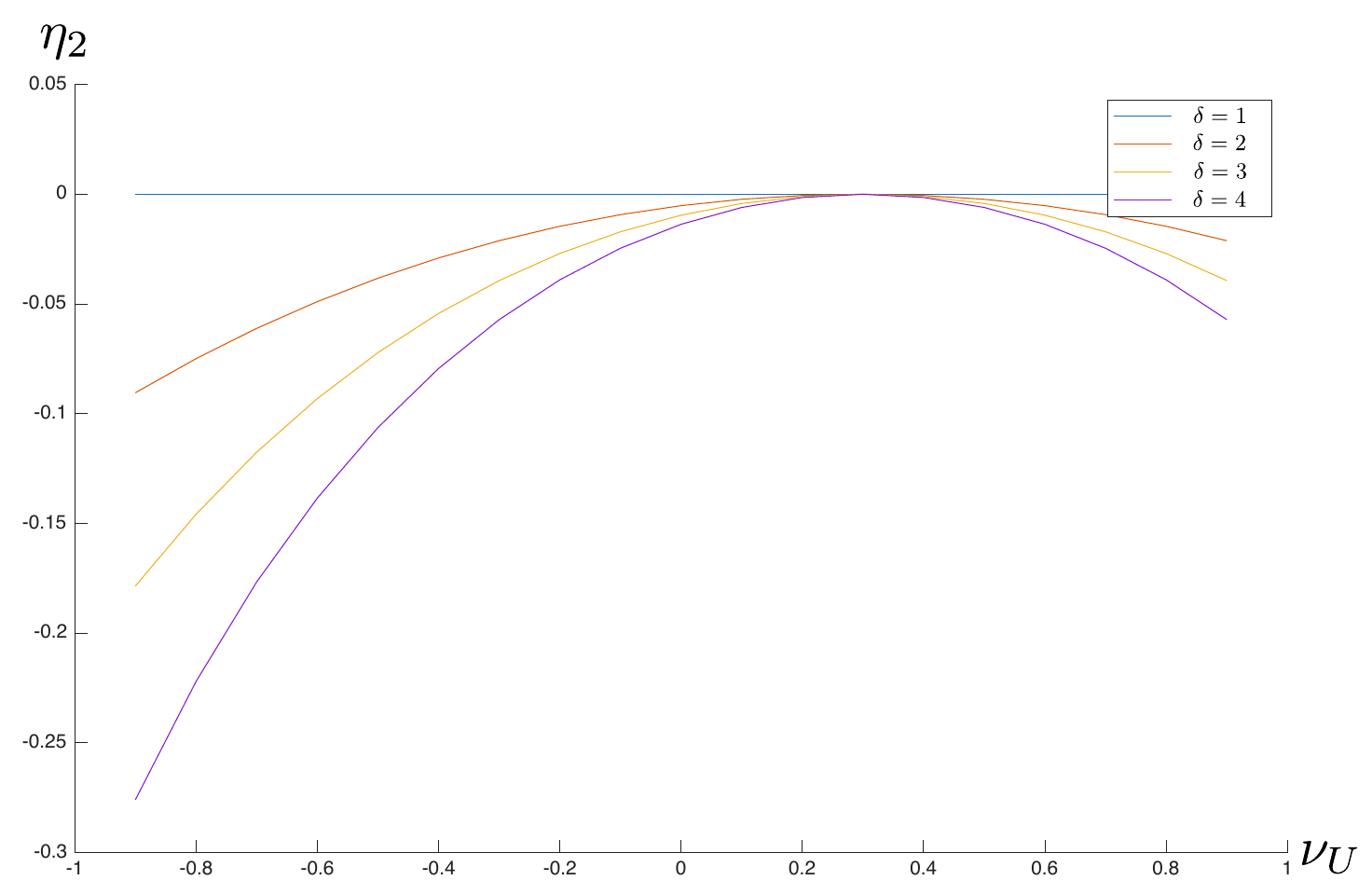} 
\caption{Percentage contribution of the transverse cross stiffness $\eta_{2}$ as a function of the upper-side Poisson's ratio $\nu_U$: (left) for fixed $a=1$, $\delta=4$, $\nu_L=0.3$, and varying $\epsilon=0.2,0.4,0.6,0.8$; (right) for fixed $a=1$, $\epsilon =0.5$, $\nu_L=0.3$, and varying $\delta=1,2,3,4$.The remaining parameters are: $\epsilon_{1L}=2\epsilon$, $\epsilon_{1U}=2$, $\nu_{1L}=0.3$, $\nu_{1U}=0.4$.}
\label{fig:percentage2}
\end{figure}

As demonstrated in Fig.~\ref{fig:percentage1}, the transverse bending percentage contribution $\eta_{22}$ can reach $10\%$--$15\%$ in slender sections, proving to be an essential correction factor (cf. \cite{le2020asymptotically,le2025asymptotically}). This effect vanishes only when the out-of-plane Poisson's ratio $\nu$ is constant, consistent with Berdichevsky's fundamental theorem \cite{berdichevskii1981energy}. Similarly, Fig.~\ref{fig:percentage2} illustrates that the cross-stiffness correction $\eta_2$ is substantial even for square sections and can be negative depending on the gradation index $\delta$, highlighting the complex energetic landscape of transversally isotropic FG-rods.

\begin{figure}[!htb]
\centering
\includegraphics[width=0.49\textwidth]{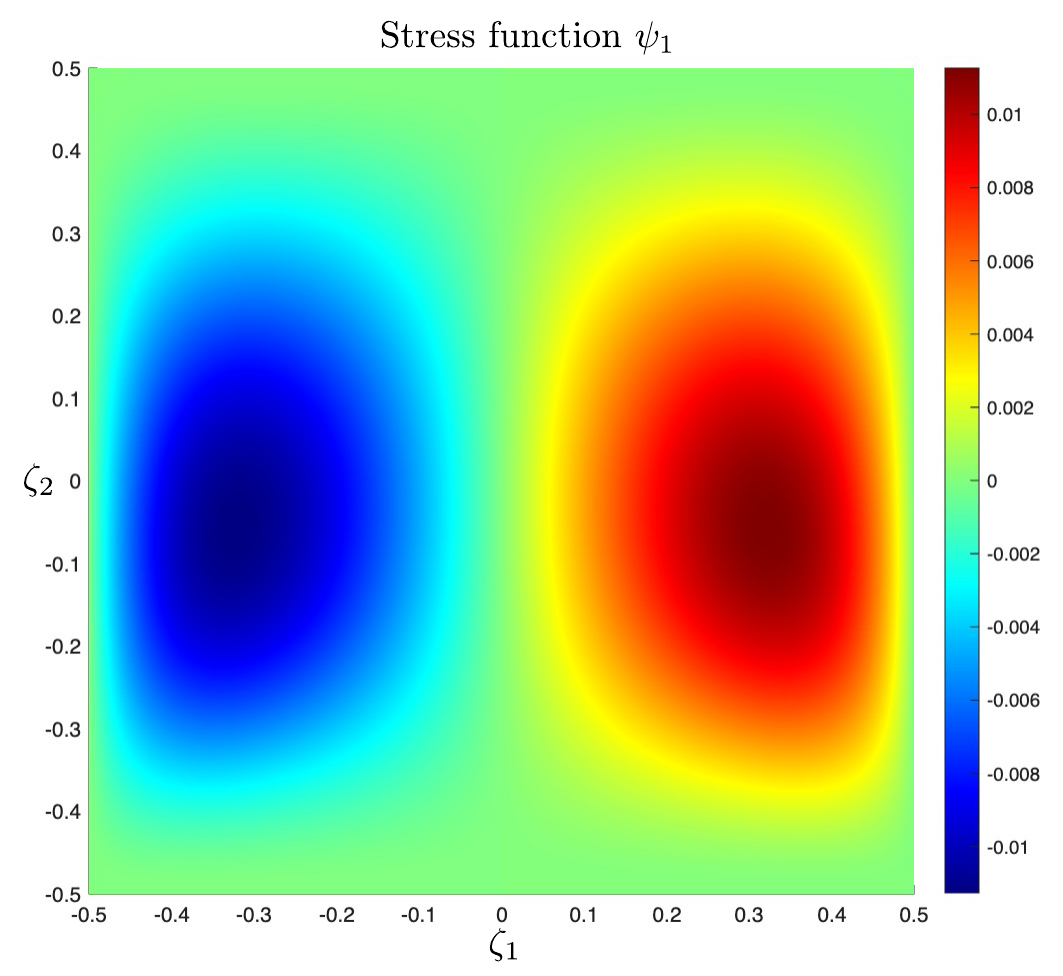}
\includegraphics[width=0.49\textwidth]{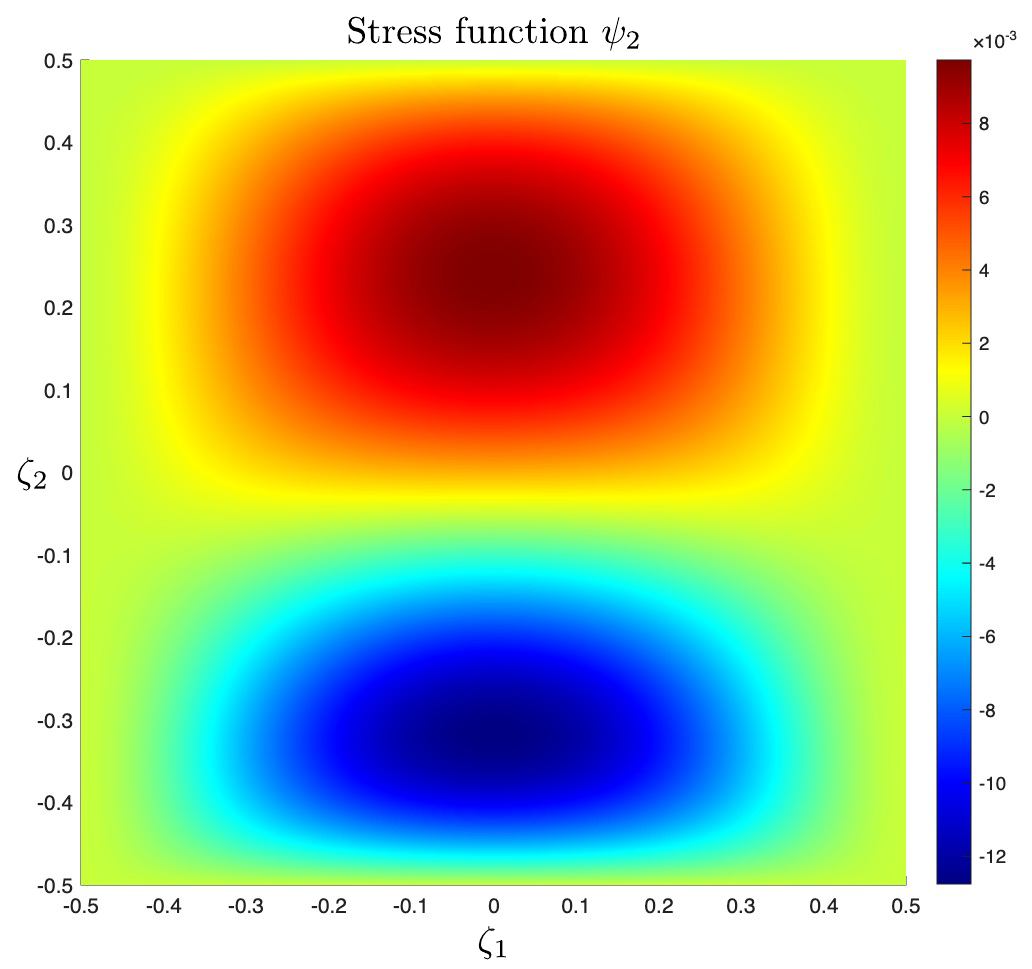} 
\caption{Contour distributions of the plane strain stress functions $\psi_1$ (left) and $\psi_{2}$ (right) for a transversally isotropic FG-rod ($a=1$, $\delta=4$, $\epsilon=0.4$, $\nu_L=0.3$, $\nu_R=0.8$). The non-zero fields illustrate the transverse coupling induced by the mismatch in Poisson's ratios.}
\label{fig:StressFunction_Pl}
\end{figure}

The numerical solutions for the stress functions $\psi_1$ and $\psi_2$, corresponding to the dual plane strain problem, are presented in Fig.~\ref{fig:StressFunction_Pl}. While their magnitudes are relatively small, their non-vanishing nature is critical; it signifies the existence of transverse stress redistributions that are entirely neglected in classical 1D models. These functions determine the positive energetic correction $\Phi_\perp$, which accounts for the cross-sectional constraints and material gradation. Consequently, the reconstructed local stress field provides a more accurate representation of the 3D state than the predictions of the naive rod theory.

\subsection{Numerical results: rhombohedral symmetry}

We now extend our analysis to functionally graded rods possessing rhombohedral symmetry. In Voigt notation, the constitutive behavior of such materials is characterized by a $6\times 6$ stiffness matrix $[c_{\mathfrak{mn}}]$ with six independent elastic constants:
\begin{equation}
\label{eq:132}
[c_{\mathfrak{mn}}]=\begin{pmatrix}
c_{11} & c_{12} & c_{13} & c_{14} & 0 & 0 \\
c_{12} & c_{11} & c_{13} & -c_{14} & 0 & 0 \\
c_{13} & c_{13} & c_{33} & 0 & 0 & 0 \\
c_{14} & -c_{14} & 0 & c_{44} & 0 & 0 \\
0 & 0 & 0 & 0 & c_{44} & c_{14} \\
0 & 0 & 0 & 0 & c_{14} & \frac{1}{2}(c_{11}-c_{12})
\end{pmatrix} .
\end{equation}
A distinctive feature of rhombohedral symmetry is the $c_{14}$ term, which induces a direct coupling between normal and shear components. Consequently, unlike the transversely isotropic case, the cross-sectional problem cannot be partitioned into independent anti-plane and plane-strain problems. Applying the asymptotic reduction logic established in Section 3, the 2D elastic compliances are derived as follows:
\begin{align}
\bar{s}_{\alpha \beta} &= \frac{1}{\bar{c}_{44}}\delta_{\alpha \beta}, \quad \bar{c}_{44}=c_{44}-\frac{2c_{14}^2}{c_{11}-c_{12}}, \label{eq:133} \\
\bar{s}_{\alpha \beta \gamma \delta } &= \bar{s}_1 \delta_{\alpha \beta}\delta_{\gamma \delta }+\bar{s}_2 (\delta_{\alpha \gamma}\delta_{\beta \delta }+\delta_{\alpha \delta}\delta_{\beta \gamma }), \label{eq:134}
\end{align}
where the compliance parameters $\bar{s}_1$ and $\bar{s}_2$ incorporate the rhombohedral coupling factor $r = c_{14}/(c_{11}-c_{12})$:
\begin{equation}
\label{eq:135}
\bar{s}_1=\frac{-c_{12}}{c_{11}^2-c_{12}^2}-\frac{r^2}{\bar{c}_{44}},\quad \bar{s}_2=\frac{1}{2(c_{11}-c_{12})}+\frac{r^2}{\bar{c}_{44}}.
\end{equation}
The interaction between the fields is governed by the non-vanishing third-rank compliance components:
\begin{equation}
\label{eq:136}
\bar{s}_{112}=\bar{s}_{121}=\bar{s}_{211}=-\frac{r}{\bar{c}_{44}}, \quad \bar{s}_{222}=\frac{r}{\bar{c}_{44}}.
\end{equation}

The remaining 2D moduli, specifically the effective longitudinal Young's modulus $E$ and the out-of-plane Poisson-like ratio $\nu$, are expressed in terms of the stiffness components by:
\begin{equation}
\label{eq:137_new}
E=c_{33}-\frac{2c_{13}^2}{c_{11}+c_{12}}, \quad \nu=\frac{c_{13}}{c_{11}+c_{12}}.
\end{equation}
The parameters ($E, c_{11}, c_{12}, \bar{c}_{44}, r, \nu$) fully define the 2D material response for rhombohedral FG-rods. For a rectangular cross-section with aspect ratio $a$, we adopt a power-law distribution across the thickness $\zeta_2$ following the form established in \eqref{eq:103-107}. The dual $\mathbf{c}$-matrix for the coupled problem is constructed as:
\begin{equation}
\label{eq:145}
\mathbf{c}^*=\begin{pmatrix}
\bar{s}_1+2\bar{s}_2 & 0 & 0 & \bar{s}_{1} & r/\bar{c}_{44} & 0 \\
0 & \bar{s}_{2}+\theta & \bar{s}_{2}-\theta & 0 & 0 & -r/\bar{c}_{44} \\
0 & \bar{s}_{2}-\theta & \bar{s}_{2}+\theta & 0 & 0 & -r/\bar{c}_{44} \\
\bar{s}_{1} & 0 & 0 & \bar{s}_1+2\bar{s}_2 & -r/\bar{c}_{44} & 0 \\
r/\bar{c}_{44} & 0 & 0 & -r/\bar{c}_{44} & 1/\bar{c}_{44} & 0 \\
0 & -r/\bar{c}_{44} & -r/\bar{c}_{44} & 0 & 0 & 1/\bar{c}_{44} 
\end{pmatrix},
\end{equation}
where $\theta$ is a large penalty parameter to enforce the symmetry constraint.

We consider an FG mixture of Barium Titanate ($\text{BaTiO}_3$) and Corundum ($\text{Al}_2\text{O}_3$), with properties listed in Table~\ref{tab:1}. To ensure energy positive-definiteness while maintaining high contrast in Poisson's ratios, a stabilized value of $c_{14} = -60 \text{ GPa}$ is utilized.

\begin{table}[htp]
    \centering
    \begin{tabular}{*{7}{p{1.2cm}}} 
        \toprule
 & $c_{11}$ & $c_{33}$ & $c_{12}$ & $c_{13}$ & $c_{44}$ & $c_{14}$ \\
        \midrule
 BaTiO$_3$  & 193 & 244 & 130 & 153 & 120 & -60 \\
        \midrule
 Al$_2$O$_3$ & 497 & 498 & 164 & 110 & 147 & -20 \\
        \bottomrule
    \end{tabular}
    \caption{Elastic moduli of phases (in GPa)}
    \label{tab:1}
\end{table}

\begin{figure}[!htb]
\centering
\includegraphics[width=0.6\textwidth]{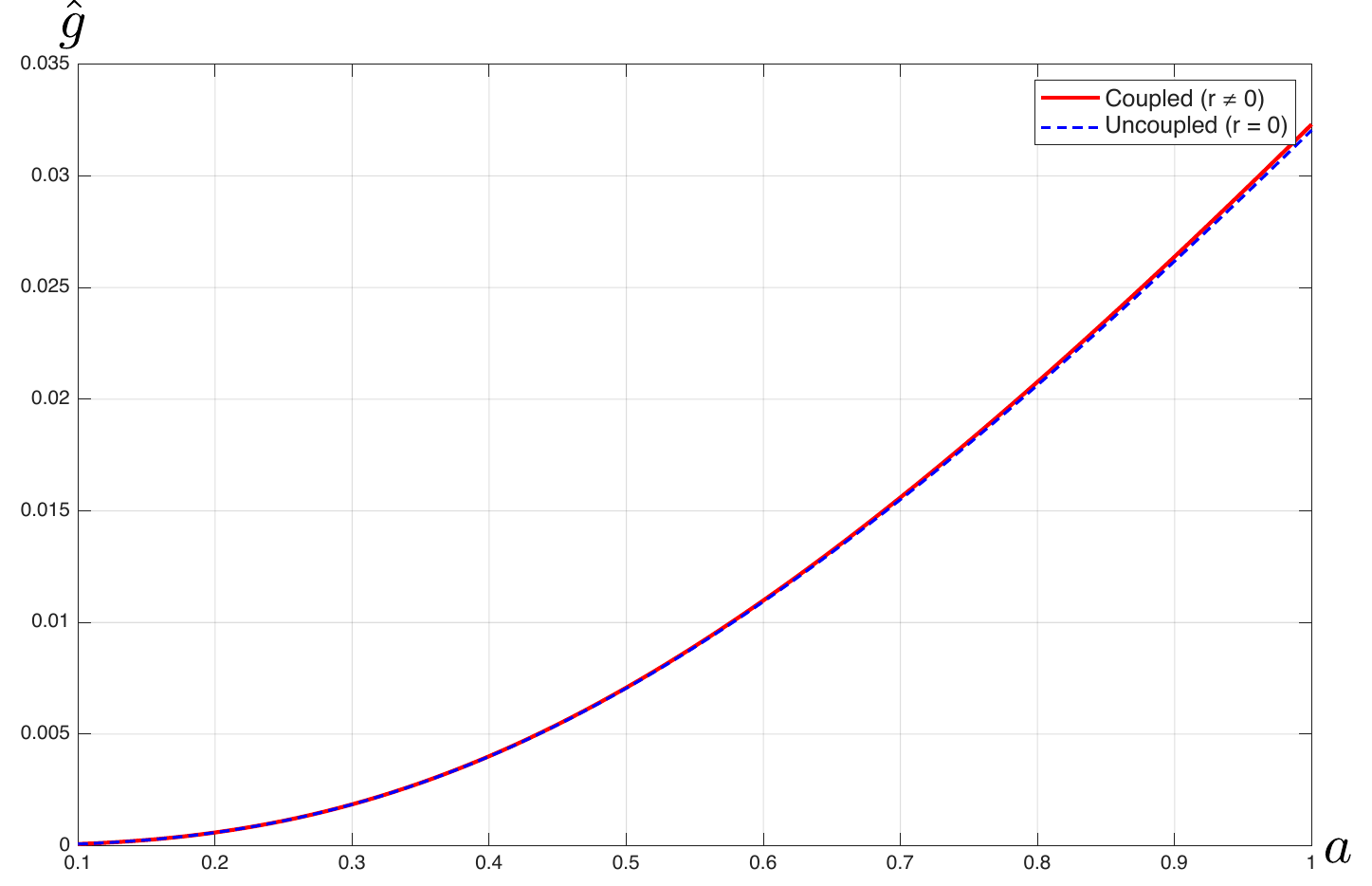}
\caption{Normalized torsional stiffness $\hat{g}$ as a function of the aspect ratio $a$ for FG-rod Barium Titanate/Corundum with $\delta=4$.}
\label{fig:5}
\end{figure}

The normalized torsional stiffness $\hat{g}$ for this coupled system is presented in Fig.~\ref{fig:5}. To isolate the influence of rhombohedral coupling, we compare these results against a case where $r=0$. The data clearly illustrates that the intrinsic coupling ($r \neq 0$) enhances the torsional resistance. Furthermore, the transverse energy density provides a $0.17\%$ correction to the bending stiffness for slender sections ($a=10$). It can be shown that, for FG-rods with higher contrast in Poisson's ratios the corrections could reach 7-8$\%$. While these corrections and the cross-stiffnesses $D, D_\alpha$ are relatively small for this material pair, the results demonstrate that the proposed framework rigorously captures the complex 3D interactions inherent to low-symmetry anisotropic FGMs.

\begin{figure}[!htb]
\centering
\includegraphics[width=0.49\textwidth]{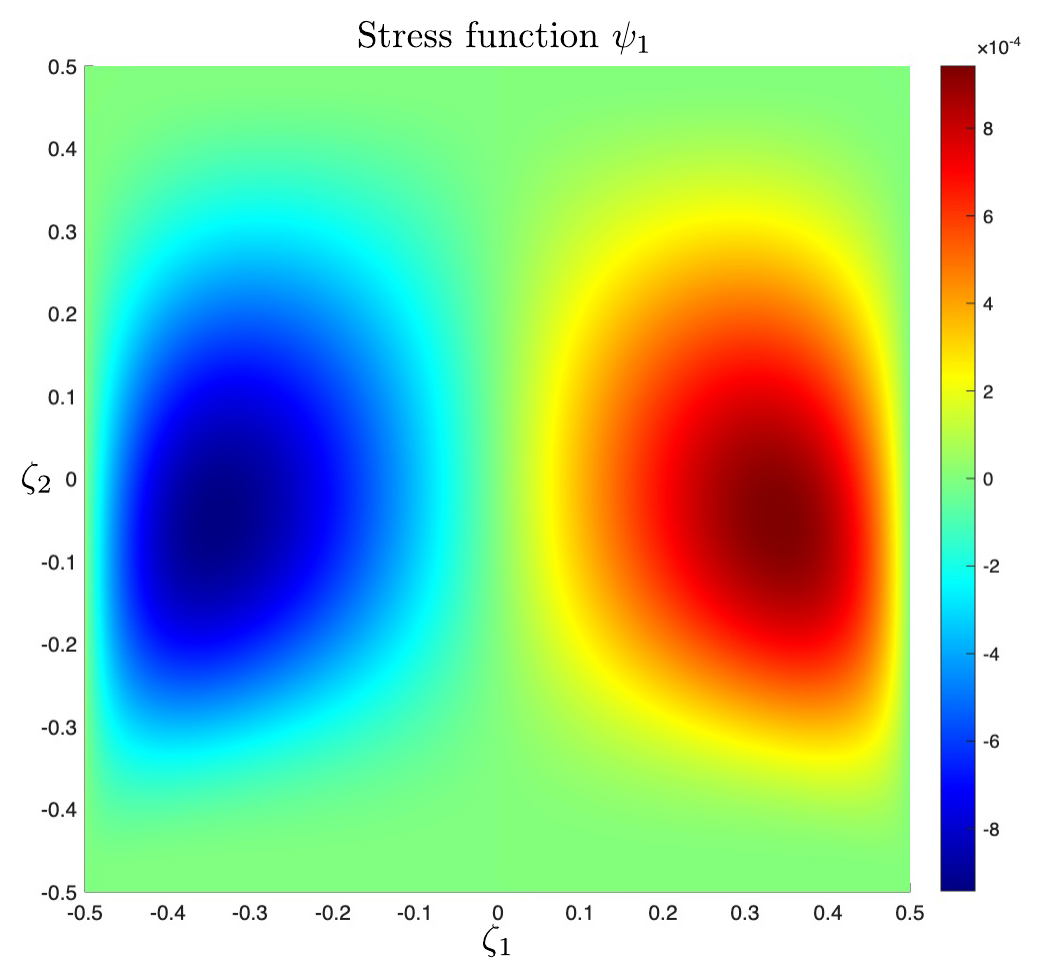}
\includegraphics[width=0.49\textwidth]{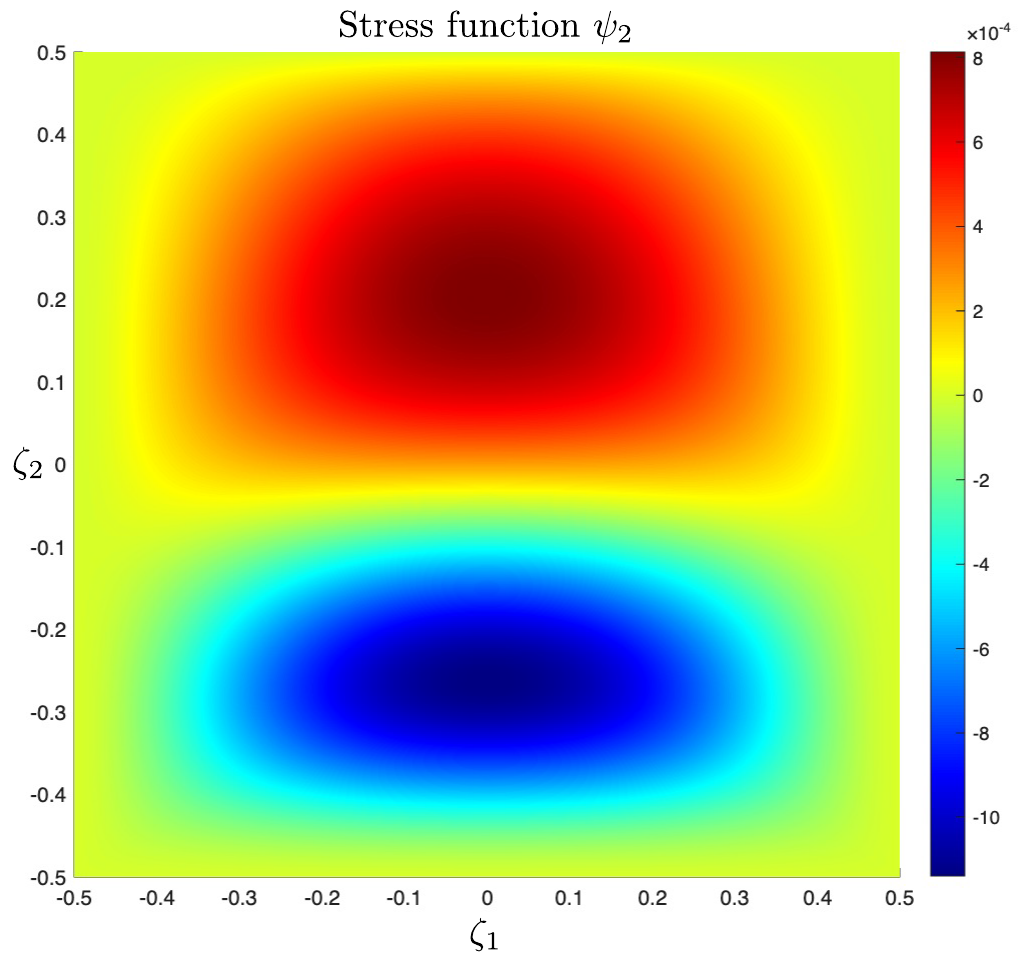} 
\caption{Contour plots of the coupled stress functions $\psi_1$ (left) and $\psi_{2}$ (right) for a rhombohedral Barium Titanate/Corundum FG-rod ($a=1$, $\delta=4$). The complex distributions reflect the constitutive coupling characteristic of rhombohedral anisotropy.}
\label{fig:StressFunction_Pl_Rhombo}
\end{figure}

The stress functions $\psi_1$ and $\psi_2$ for the rhombohedral FG-rod, obtained by solving the fully coupled dual problem, are displayed in Fig.~\ref{fig:StressFunction_Pl_Rhombo}. In contrast to higher-symmetry cases, these distributions capture the intricate coupling between extension, bending, and shear induced by the $c_{14}$ stiffness component. The highly non-trivial morphology of these contours highlights the significant deviation of the local 3D stress state from the simplified predictions of classical rod theories. These results underscore the necessity of the variational-asymptotic approach for accurately capturing the energetic landscape of low-symmetry functionally graded materials.

\section{1D theory and restoration of the 3D stress and strain fields}

The ultimate utility of the variational-asymptotic method (VAM) lies not only in the reduction of dimensionality but also in the ability to reconstruct the local three-dimensional state from the global one-dimensional solution. This section details the governing 1D equations and provides a systematic procedure for the restoration of 3D displacement, strain, and stress fields, a feature that allows for detailed local analysis within a computationally efficient framework.

\subsection{One-dimensional governing equations}

With the dual cross-sectional problems \eqref{eq:41} and \eqref{eq:63} resolved, we can now establish the effective one-dimensional (1D) framework for functionally graded rods. To maintain rigor while working with 3D curvilinear coordinates where the metric tensor is not identity, we employ standard tensor notation to distinguish between co- and contravariant components.

The dimensional reduction is performed by substituting the displacement field \eqref{eq:29}--\eqref{eq:30}--utilizing the minimizers $y_\alpha (\zeta^\alpha,x,t)$ and $y(\zeta^\alpha,x,t)$ of \eqref{eq:41}--into the original 3D action functional \eqref{eq:2}. By integrating over the cross-sectional domain and maintaining only the asymptotically principal terms, we derive the 1D action functional:
\begin{equation}
\label{eq:147}
J[u(x,t),u_\alpha(x,t),\phi(x,t)]=\int_{t_0}^{t_1}\int_0^L [\Theta(\dot{u},\dot{u}_\alpha,\dot{\phi})-\Phi (\gamma,\Omega_\alpha,\Omega)] \dd{x}\dd{t}.
\end{equation}
As established in Section 4, the 1D strain energy density follows Eq.~\eqref{eq:44}, with the kinematic measures defined in \eqref{eq:33}--\eqref{eq:35}. Utilizing the results from \eqref{eq:46} and \eqref{eq:96}, the energy density is expressed as:
\begin{equation}
\label{eq:148}
\Phi (\gamma,\Omega_\alpha,\Omega)=\frac{1}{2}[\bar{E}\gamma^2+B_{\alpha \beta}\Omega^\alpha \Omega^\beta+C\Omega^2+2\gamma (B_\alpha \Omega^\alpha +D\Omega )+2D_\alpha \Omega^\alpha \Omega ],
\end{equation}
where the effective stiffnesses are summed from the longitudinal and transverse contributions:
\begin{equation}
\label{eq:149}
\bar{E}=E_\parallel +E_\perp ,\quad B_{\alpha \beta}=B_{\alpha \beta \parallel }+B_{\alpha \beta \perp },\quad B_{\alpha }=B_{\alpha \parallel }+B_{\alpha \perp }.
\end{equation}
The kinetic energy density, following Eq.~\eqref{eq:46}, is cast in the form:
\begin{equation}
\label{eq:150}
\Theta(\dot{u},\dot{u}_\alpha,\dot{\phi})=\frac{1}{2}[ \bar{\rho} (\dot{u}^2+\dot{u}_\alpha \dot{u}^\alpha )-2\bar{\rho}_\alpha \dot{u}^\alpha \dot{\phi}+\bar{\rho}_r \dot{\phi}^2 ],
\end{equation}
where the effective inertial properties are:
\begin{equation}
\label{eq:151}
\bar{\rho}=h^2\langle \rho \rangle, \quad \bar{\rho}_\alpha =h^3 \langle \rho e_{\alpha \beta}\zeta^\beta \rangle, \quad \bar{\rho}_r=h^4 \langle \rho \zeta _\alpha \zeta^\alpha \rangle .
\end{equation}
Here, $\bar{\rho}$ represents the effective mass per unit length, $\bar{\rho}_\alpha$ denotes the mass coupling inertia, and $\bar{\rho}_r$ is the effective rotary inertia.

The dynamic behavior of the FG-rod is governed by a variational principle: the actual average displacements $\check{u}, \check{u}_\alpha$ and the twist angle $\check{\phi}$ correspond to the stationary point of the action functional \eqref{eq:147} among all admissible fields. This stationarity condition leads to the following set of 1D equations of motion:
\begin{align}
\bar{\rho }\ddot{u}&=T^\prime +M^\prime _\alpha \omega ^\alpha  
-e^{\alpha \beta }M_\beta \varpi \omega _\alpha ,
\label{eq:152} \\
\bar{\rho }\ddot{u}_\alpha -\bar{\rho}_\alpha \ddot{\phi} &=-T\omega _\alpha
+M^{\prime \prime }_\alpha    
-e_{\alpha .}^{. \beta }[M_\beta ^\prime \varpi
+(M_\beta \varpi )^\prime 
-(M\omega _\beta )^\prime ]
\notag \\
&-(M_\alpha \varpi -M\omega _\alpha )\varpi ,
\label{eq:153} \\
\bar{\rho }_\phi \ddot{\phi } -\bar{\rho}_\alpha \ddot{u}^\alpha &=
M^\prime -e^{\alpha \beta }M_\alpha  \omega _\beta ,
\label{eq:154}
\end{align} 
supplemented by initial and boundary conditions. The internal force and moments are defined through the constitutive relations:
\begin{align}
&T=\frac{\partial \Phi}{\partial \gamma}=\bar{E}\gamma +B_{\alpha}\Omega^\alpha +D\Omega, \label{eq:155} \\
&M_\alpha =\frac{\partial \Phi}{\partial \Omega_\alpha}=B_{\alpha \beta }\Omega^\beta +B_{\alpha}\gamma +D_\alpha \Omega, \label{eq:156} \\
&M=\frac{\partial \Phi}{\partial \Omega}=C \Omega +D\gamma +D_\alpha \Omega^\alpha. \label{eq:157}
\end{align}

To account for external traction $\vb*{\tau}$ acting on the lateral boundary, we modify the action functional \eqref{eq:2} as follows:
\begin{equation}
\label{eq:158}
I=\int_{t_0}^{t_1}\int_0^L \Bigl[ \int_{\mathcal{S}}
[T(x^\alpha ,\dot{\mathbf{w}})-W(x^\alpha ,\boldsymbol{\varepsilon })]\sqrt{g} \dd{a}+\int_{\partial \mathcal{S}} \vb*{\tau }\cdot \mathbf{w} \dd{s} \Bigr] \dd{x}\dd{t}.
\end{equation}
Assuming the traction magnitude satisfies the asymptotic scaling to maintain the small strain $\varepsilon$ in the limit $h\to 0$, $\mathbf{\tau} \sim O( E \varepsilon (h/l+h/R) )$, the resulting corrections to the warping functions are negligible in the first-order approximation. Integration yields the augmented 1D functional:
\begin{equation}
\label{eq:160}
J=\int_{t_0}^{t_1}\int_0^L [\Theta(\dot{u},\dot{u}_\alpha,\dot{\phi})-\Phi (\gamma,\Omega_\alpha,\Omega)+F^\alpha u_\alpha +Fu+Q\phi ] \dd{x}\dd{t},
\end{equation}
where the distributed loads $F^i$ and twisting moment $Q$ are defined by:
\begin{equation}
\label{eq:161}
F^i=h \int_{\partial \bar{\mathcal{S}}} \tau^i \dd{s}, \quad Q=-h^2 \int_{\partial \bar{\mathcal{S}}} e_{\alpha \beta}\tau^\alpha \zeta^\beta \dd{s}.
\end{equation}
The corresponding governing equations for the loaded rod become:
\begin{align}
\bar{\rho }\ddot{u}&=T^\prime +M^\prime _\alpha \omega ^\alpha  
-e^{\alpha \beta }M_\beta \varpi \omega _\alpha +F,
\label{eq:162} \\
\bar{\rho }\ddot{u}_\alpha -\bar{\rho}_\alpha \ddot{\phi} &=-T\omega _\alpha
+M^{\prime \prime }_\alpha    
-e_{\alpha .}^{.\beta }[M_\beta ^\prime \varpi
+(M_\beta \varpi )^\prime 
-(M\omega _\beta )^\prime ]
\notag \\
&-(M_\alpha \varpi -M\omega _\alpha )\varpi +F_\alpha ,
\label{eq:163} \\
\bar{\rho }_\phi \ddot{\phi } -\bar{\rho}_\alpha \ddot{u}^\alpha &=
M^\prime -e^{\alpha \beta }M_\alpha  \omega _\beta +Q.
\label{eq:164}
\end{align}

\subsection{Restoration of 3D stress and strain fields}

Following the determination of the 1D kinematic fields--the displacements and twist angle--from the rod theory, a systematic procedure is established to reconstruct the 3D response. In the first-order approximation, the 3D strain field is restored via the kinematic relations \eqref{eq:9}--\eqref{eq:11}. Here, the displacement fields $w_\alpha = \check{w}_\alpha$ and $w = \check{w}$ are evaluated using the decompositions \eqref{eq:29}--\eqref{eq:30}, incorporating the optimal warping functions $y_\alpha = \check{y}_\alpha$ and $y = \check{y}$ obtained from the cross-sectional analysis \eqref{eq:41}.

The 3D stress components are then recovered through the localized constitutive relations:
\begin{align}
\sigma^{\alpha \beta}&=\pdv{W}{\varepsilon_{\alpha \beta}}=\pdv{W_\perp}{\varepsilon_{\alpha \beta}}= c^{\alpha \beta \gamma \delta}\gamma_{\gamma \delta}+2c^{\alpha \beta 3\gamma}\gamma_{\gamma}, \label{eq:165}
\\
\sigma^{3\alpha }&=\pdv{W}{(2\varepsilon_{3\alpha })}=\pdv{W_\perp}{(2\varepsilon_{3\alpha })}=c^{3\alpha 3\beta} \gamma_{\beta}+2c^{3\alpha \gamma \delta}\gamma_{\gamma \delta},
\label{eq:166}
\\
\sigma^{33}&=\pdv{W}{\varepsilon_{33}}=\pdv{W_\parallel}{\varepsilon_{33}}+\pdv{W_\perp}{\varepsilon_{33}}=\sigma^{33}_{\parallel} +\sigma^{33}_{\perp} \notag
\\
&=E\varepsilon_{33} +(c^{\alpha \beta \gamma \delta}\gamma_{\gamma \delta}+2c^{\alpha \beta 3\gamma}\gamma_{\gamma})r_{\alpha \beta}+(c^{3\alpha 3\beta} \gamma_{\beta}+2c^{3\alpha \gamma \delta}\gamma_{\gamma \delta})r_\alpha ,
\label{eq:167}
\end{align} 
where the strain measures $\gamma_{\alpha \beta}$ and $\gamma_\alpha$ are defined in \eqref{eq:42}--\eqref{eq:43}. This recovered stress field identifies with the 3D equilibrium requirements and identically satisfies the traction-free boundary conditions on the rod's lateral surface:
\begin{equation}
\label{eq:168}
\sigma^{\alpha \beta}n_\beta =0, \quad \sigma^{3\alpha }n_\alpha =0 \text{ at } \partial \mathcal{S}.
\end{equation}

The macroscopic 1D stress resultants are linked to the local 3D stress components through the following integral identities:
\begin{align}
\label{eq:169}
&T=h^2\langle \kappa \tau^{33} \rangle ,\quad V^\alpha =h^2\langle \tau^{3\alpha} +h\tau^{33}\varpi e_{\rho .}^{. \alpha} \zeta^\rho\rangle, 
\\
&M^\alpha =h^3\langle \kappa \tau^{33} \zeta^\alpha \rangle , \quad M=h^3 \langle e_{\alpha \beta} \zeta^\alpha (\tau^{3\beta }+h\tau^{33}\varpi e_{\rho .}^{. \beta} \zeta^\rho)\rangle ,
\label{eq:170}
\end{align}
where $\tau^{ij}=\kappa \sigma^{ij}$ denote the weighted stresses. These definitions, which include the shear forces $V^\alpha$ as well, account for the rod's initial curvature and twist. While the geometric factor $\kappa$ and initial twist $\varpi$ can often be simplified in first-order approximations, we maintain their exact definitions in \eqref{eq:169}--\eqref{eq:170} to ensure strict compatibility with the 3D equilibrium equations as will be seen later.

To verify the validity of these relationships, we represent the minimized transverse energy $\Phi_\perp$ in an expanded form by utilizing \eqref{eq:16} and \eqref{eq:42}--\eqref{eq:43}:
\begin{multline}
\label{eq:171}
\Phi_\perp =\frac{1}{2}h^2 \langle c^{\alpha \beta \gamma \delta }
(\check{y}_{(\alpha ,\beta )}+
r_{\alpha \beta }(\gamma +h\Omega _\sigma \zeta ^\sigma )) (\check{y}_{(\gamma ,\delta )}+
r_{\gamma \delta}(\gamma +h\Omega _\nu \zeta ^\nu )) 
\\
+c^{3\alpha 3\beta }(\check{y}_{,\alpha }-he_{\alpha \gamma}\Omega \zeta 
^\gamma +r_{\alpha }(\gamma +h\Omega _\sigma \zeta ^\sigma ))(\check{y}_{,\beta }-he_{\beta \delta}\Omega \zeta 
^\delta +r_{\beta }(\gamma +h\Omega _\nu \zeta ^\nu ))
\\
+2c^{\alpha \beta 3\gamma }(\check{y}_{(\alpha ,\beta )}+
r_{\alpha \beta }(\gamma +h\Omega _\sigma \zeta ^\sigma ))
(\check{y}_{,\gamma }-he_{\gamma \delta}\Omega \zeta 
^\delta +r_{\gamma }(\gamma +h\Omega _\nu \zeta ^\nu )) \rangle .
\end{multline}
By virtue of the Euler--Lagrange equations for the warping minimizers, the first variation of the transverse energy vanishes, leading to the identity:
\begin{multline}
\label{eq:172}
\langle c^{\alpha \beta \gamma \delta }
(\check{y}_{(\gamma ,\delta )}+
r_{\gamma \delta}(\gamma +h\Omega _\sigma \zeta ^\sigma )) \check{y}_{(\alpha ,\beta )}
+c^{3\alpha 3\beta }(\check{y}_{,\beta }-he_{\beta \sigma}\Omega \zeta 
^\sigma +r_{\beta }(\gamma +h\Omega _\sigma \zeta ^\sigma ))\check{y}_{,\alpha }
\\
+2c^{\alpha \beta 3\gamma }(\check{y}_{(\alpha ,\beta )}+
r_{\alpha \beta }(\gamma +h\Omega _\sigma \zeta ^\sigma ))
\check{y}_{,\gamma }
\\
+2c^{\alpha \beta 3\gamma }\check{y}_{(\alpha ,\beta )}
(\check{y}_{,\gamma }-he_{\gamma \delta}\Omega \zeta 
^\delta +r_{\gamma }(\gamma +h\Omega _\nu \zeta ^\nu )) \rangle =0.
\end{multline}
Applying \eqref{eq:166}--\eqref{eq:167}, the remaining terms reduce to an energetic equivalence:
\begin{equation}
\label{eq:173}
\Phi_\perp = \frac{1}{2}h^2 \langle \sigma^{33}_{\perp} (\gamma +h\Omega _\sigma \zeta ^\sigma )-\sigma^{3\alpha }he_{\alpha \beta}\Omega \zeta^\beta \rangle .
\end{equation}

For specific deformation modes, such as pure torsion ($\gamma = 0, \Omega_\alpha = 0$), the relationship \eqref{eq:173} equates the 1D twisting moment to the 3D shear stress integral leading to the last equation of \eqref{eq:170}. Similar considerations confirm also the first equations of \eqref{eq:169} and \eqref{eq:170}, where we should also take the contribution of $\sigma^{33}_{\parallel }=E\varepsilon_{33}$ into account. For a general coupled state, we represent the strain state $\varepsilon=(\varepsilon_{33},\varepsilon_{3\alpha})$ as a superposition of extension $\varepsilon^\prime=(\gamma,0)$, bending $\varepsilon^{\prime \prime}=(h\Omega_\alpha \zeta^\alpha ,0)$, and torsion $\varepsilon^{\prime \prime \prime}=(0,-he_{\alpha \beta}\Omega \zeta^\beta )$. Utilizing the scalar product $\langle \sigma \varepsilon \rangle = \langle \sigma^{33}_{\perp} \varepsilon_{33} + \sigma^{3\alpha }2\varepsilon_{3\alpha} \rangle$, the transverse energy density can be decomposed based on the independent 1D strain measures:
\begin{multline}
\label{eq:177}
\Phi_\perp =\frac{1}{2}[(E_\perp \gamma +B_{\alpha \perp}\Omega ^\alpha +D\Omega)\gamma+(B_{\alpha \beta \perp}\Omega^\beta +B_{\alpha \perp}\gamma +D_\alpha \Omega)\Omega^\alpha 
\\
+(C\Omega +D\gamma +D_\alpha \Omega^\alpha)\Omega],
\end{multline} 
This decomposition yields:
\begin{align}
h^2\langle \sigma^{33}_{\perp} \rangle &=E_\perp \gamma +B_{\alpha \perp}\Omega ^\alpha +D\Omega , \label{eq:178} \\
h^3\langle \sigma^{33}_{\perp} \zeta _\alpha \rangle &=B_{\alpha \beta \perp}\Omega^\beta +B_{\alpha \perp}\gamma +D_\alpha \Omega , \label{eq:179} \\
h^3 \langle e_{\alpha \beta}\zeta^\alpha \sigma^{3\beta }\rangle &=C\Omega +D\gamma +D_\alpha \Omega^\alpha . \label{eq:180}
\end{align}
Finally, by incorporating the longitudinal contribution $\sigma^{33}_{\parallel} = E\varepsilon_{33}$, the complete definitions of the 1D force resultants \eqref{eq:169}--\eqref{eq:170} are rigorously recovered from the local 3D stress state.

\section{Error estimation of the one-dimensional theory}

In this section, we utilize the Prager--Synge identity \cite{prager1947approximations} to establish a rigorous error estimate for the 1D functionally graded rod theory developed in the preceding sections. This methodology allows us to quantify the deviation of the reduced model from the exact three-dimensional elasticity solution.

\subsection{The Prager-Synge framework}
We consider the linear space of stress fields $\boldsymbol{\sigma}(\mathbf{x})$ equipped with the energetic norm:
\begin{equation}
\|\boldsymbol{\sigma}\|^2_{L_2}=C_2[\boldsymbol{\sigma}]=\int_{\mathcal{V}} W^*(\mathbf{x},\boldsymbol{\sigma})\text{d}^3x.
\label{eq:181}
\end{equation}
Here, $W^*(\mathbf{x},\boldsymbol{\sigma})$ represents the positive-definite complementary energy density:
\begin{equation}
W^*(\mathbf{x},\boldsymbol{\sigma})=\frac{1}{2}\boldsymbol{\sigma}\boldsymbol{:}\mathbf{s}(\mathbf{x})\boldsymbol{:}\boldsymbol{\sigma},
\label{eq:182}
\end{equation}
where $\mathbf{s}$ is the fourth-rank tensor of elastic compliances. To apply the identity, we define two classes of admissible stress fields: 
\begin{enumerate}
  \item \textbf{Kinematically admissible stress fields} $\hat{\boldsymbol{\sigma}}$: These are derived from a displacement field $\hat{\mathbf{u}}$ that satisfies the kinematic compatibility relation:
\begin{equation}
\label{eq:183}
\hat{\boldsymbol{\varepsilon}} = \frac{1}{2}(\nabla \hat{\mathbf{u}}+(\nabla \hat{\mathbf{u}})^T),
\end{equation}
where the strain $\hat{\boldsymbol{\varepsilon}}$ is linked to $\hat{\boldsymbol{\sigma}}$ through the material's constitutive law.
  \item \textbf{Statically admissible stress fields} $\tilde{\boldsymbol{\sigma}}$: These fields satisfy the point-wise 3D equilibrium equations:
\begin{equation}
\text{div } \tilde{\boldsymbol{\sigma}}=\mathbf{0}, \quad \tilde{\boldsymbol{\sigma}}\cdot \mathbf{n}=\mathbf{0} \quad \text{at } \partial \mathcal{S}\times (0,L).
\label{eq:184}
\end{equation}
\end{enumerate}

The kinematic and static boundary conditions at the rod's edges must be satisfied by $\hat{\mathbf{u}}$ and $\tilde{\boldsymbol{\sigma}}$, respectively. Let $\check{\boldsymbol{\sigma}}$ denote the true stress state within the FG rod. The Prager--Synge identity\footnote{For generalizations of this identity to piezoelectricity and its application in error estimation for laminated and functionally graded shells, see \cite{le1986theory,le2016asymptotically,le2017asymptotically}.} relates the exact stress to any pair of admissible fields as follows:
\begin{equation}
\label{eq:185}
C_2[\check{\boldsymbol{\sigma}}-\frac{1}{2}(\hat{\boldsymbol{\sigma}}+\tilde{\boldsymbol{\sigma}})]=C_2[\frac{1}{2}(\hat{\boldsymbol{\sigma}}-\tilde{\boldsymbol{\sigma}})].
\end{equation}
This identity implies that if a kinematically admissible field $\hat{\boldsymbol{\sigma}}$ and a statically admissible field $\tilde{\boldsymbol{\sigma}}$ can be found such that their difference is small in the energetic norm, then their average provides a high-fidelity approximation of the true stress, with the error bounded by the distance between the two fields.

\subsection{Convergence theorem}

\textbf{Theorem.} \textit{The stress state determined by the 1D FG rod theory differs from the exact 3D stress state by quantities of order $O(h/l + h/R)$ in the $L_2$ norm.}

To prove this, we construct kinematically and statically admissible fields that remain asymptotically close to the 1D rod theory predictions.

\textbf{Kinematically admissible stress state.} We define the admissible displacement field $\hat{\mathbf{u}}$ as:
\begin{align}\label{eq:186}
&\hat{u}_\alpha =u_\alpha (x,t)-he_{\alpha \beta}\phi (x,t)\zeta ^\beta+hy_\alpha (\zeta ^\alpha,x,t), \\
&\hat{u}=u(x,t)-h e_{\alpha .}^{. \beta} \phi_{\beta} (x,t) \zeta^\alpha +hy(\zeta^\alpha,x,t),
\label{eq:187}
\end{align}
where $\phi_\alpha$ follows \eqref{eq:31}, and $u_i, \phi$ are the 1D rod solutions. The functions $y_\alpha$ and $y$ are the cross-sectional minimizers of \eqref{eq:41}. Assuming the 3D kinematic boundary conditions at the edges are compatible with this inner expansion, the strain field derived from \eqref{eq:183} is kinematically admissible. Asymptotic analysis confirms that $\hat{\boldsymbol{\varepsilon}}$ deviates from the 1D theory's strain by terms of order $O(h/l+h/R)$. Consequently, the associated stress $\hat{\boldsymbol{\sigma}}$ differs from the recovered stresses in \eqref{eq:165}--\eqref{eq:167} by the same asymptotic order. Note that the proof can be extended to irregular boundary conditions at the rod edges by considering the additional, exponentially decaying, strain field in the boundary layer, which contributes an error of order $O(h/l)$ \cite{gregory1984decaying,ladeveze1998new}.

\textbf{Statically admissible field.} The exact 3D equilibrium equations for the weighted stresses $\tau^{ij}=\kappa \sigma^{ij}$ in curvilinear coordinates are:
\begin{align}
&(\tau^{\alpha \beta})_{, \beta}+he_{\beta .}^{. \alpha}\varpi (\tau ^{3\gamma }\zeta^\beta)_{,\gamma}+h(\tau^{3\alpha })_{,x}+h\varpi e_{\beta .}^{.\alpha }\tau^{3\beta } \notag
\\
&-h\omega^\alpha \kappa\tau^{33}+h^2(\tau^{33}\varpi )_{,x}e_{\beta .}^{. \alpha }\zeta^\beta+h^2\tau^{33}\varpi^2 \zeta^\alpha=0,
\label{eq:188}
\\
&(\tau^{3\alpha }\kappa)_{,\alpha} +h(\tau^{33}\kappa)_{,x} +h\omega_\alpha \tau^{3\alpha}+h^2\tau^{33}\varpi e_{\alpha .}^{. \beta}\zeta^\alpha \omega_\beta=0,
\label{eq:189}
\end{align}
with traction-free conditions $\tau^{\alpha \beta }n_\beta =0$ and $\tau^{3\alpha }n_\alpha =0$. Integrating these equations over the cross-section confirms they reduce to the 1D equilibrium equations \eqref{eq:152}--\eqref{eq:154} when the inertial terms are neglected (static case).

We construct $\tilde{\boldsymbol{\sigma}}$ by setting the longitudinal component $\tilde{\sigma}^{33}$ to the value recovered from the 1D theory \eqref{eq:167}. The remaining components $\tilde{\sigma}^{\alpha \beta}$ and $\tilde{\sigma}^{3\alpha }$ are formulated as:
\begin{align}
\label{eq:195}
\tilde{\sigma}^{\alpha \beta}&=c^{\alpha \beta \gamma \delta}(\tilde{y}_{(\gamma ,\delta )}+
r_{\gamma \delta}(\gamma +h\Omega _\sigma \zeta ^\sigma )) \notag
\\
&+2c^{\alpha \beta 3\gamma}(\tilde{y}_{,\gamma }-he_{\gamma \nu }\Omega \zeta 
^\nu +r_{\gamma }(\gamma +h\Omega _\sigma \zeta ^\sigma )),
\\
\tilde{\sigma}^{3\alpha }&=c^{3\alpha 3\beta} (\tilde{y}_{,\beta }-he_{\beta \nu }\Omega \zeta 
^\nu +r_{\beta }(\gamma +h\Omega _\sigma \zeta ^\sigma )) \notag
\\
&+2c^{3\alpha \gamma \delta}(\tilde{y}_{(\gamma ,\delta )}+
r_{\gamma \delta}(\gamma +h\Omega _\sigma \zeta ^\sigma )),
\label{eq:196}
\end{align} 
where $\tilde{y}_i, \tilde{y}$ are determined such that the 3D equilibrium equations \eqref{eq:188}--\eqref{eq:189} are satisfied. This yields a boundary-value problem that differs from the primary cross-sectional problem only by $O(h/l + h/R)$ terms. Although boundary conditions at the rod ends are only met "on average," Saint-Venant's principle \cite{toupin1965saint,berdichevskii1974proof,ladeveze1998new} ensures that these discrepancies decay exponentially, contributing an error only of order $O(h/l)$.

Since both admissible fields $\hat{\boldsymbol{\sigma}}$ and $\tilde{\boldsymbol{\sigma}}$ deviate from the 1D theory by $O(h/l + h/R)$, the Prager--Synge identity \eqref{eq:185} guarantees that the total energetic error of the 1D rod theory is of the same order, thus concluding the proof.

\begin{figure}[htb]
    \centering
   \includegraphics[width=0.49\textwidth]{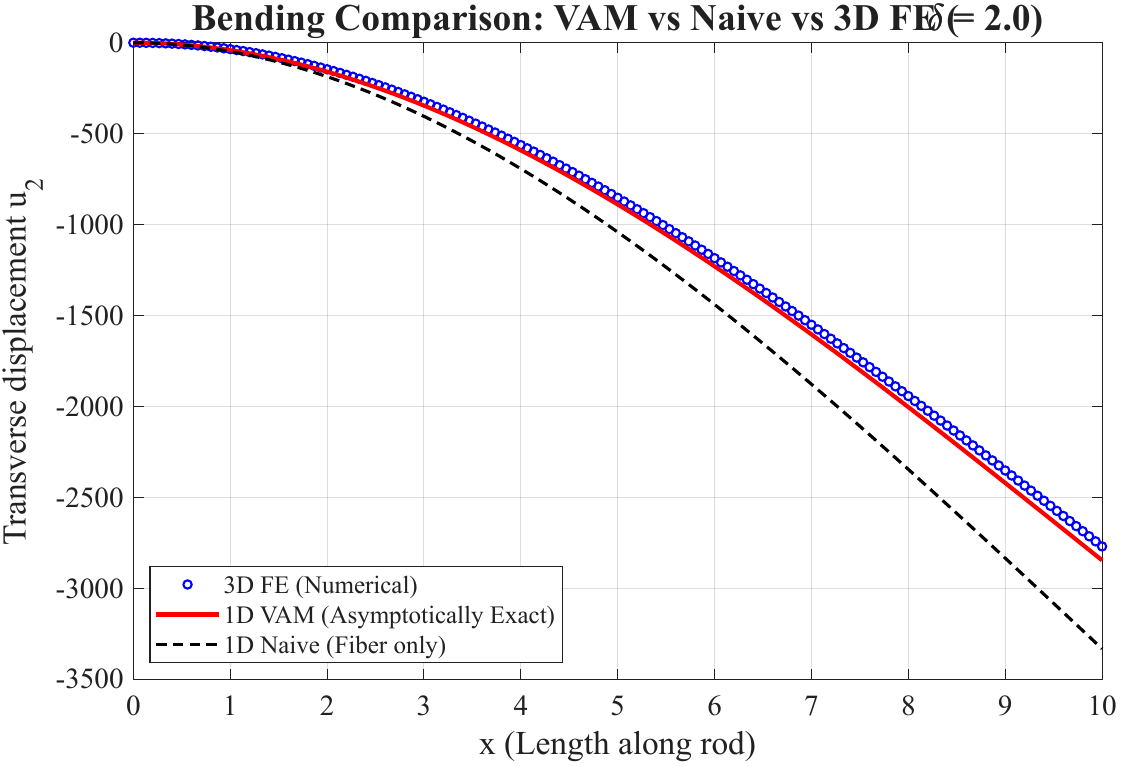}
\includegraphics[width=0.49\textwidth]{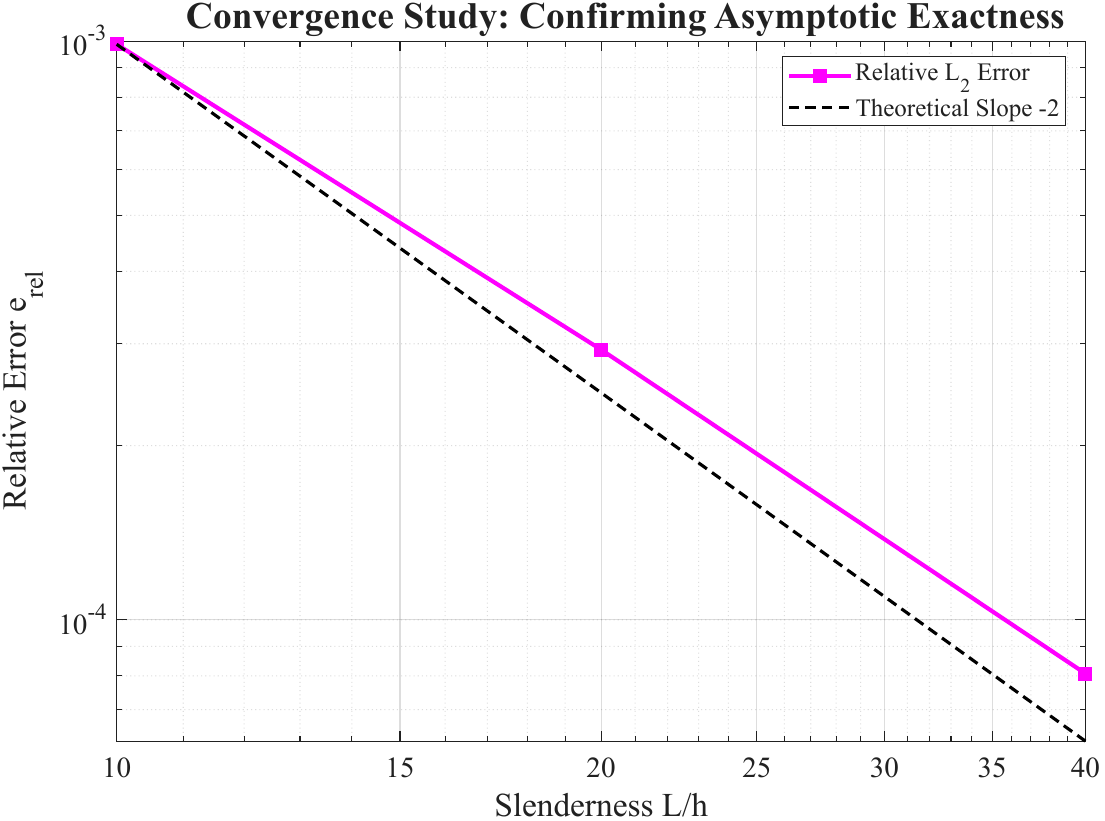}
    \caption{(left) Deflection of the straight FG beam of rectangular cross section versus its length, (right) Relative $L_2$-error versus slenderness $L/h$.}
    \label{fig:accuracy}
\end{figure}

Fig. \ref{fig:accuracy} illustrates the asymptotic accuracy of the proposed 1D FG rod theory through a benchmark problem: a transversally isotropic FG straight beam of rectangular cross-section subjected to a concentrated tip load $P_2$. The dimensionless parameters are chosen as follows: $L = 10$, $h = 1$, $a = 1.5$, $\delta = 2.0$, $P_2= 1$, $\epsilon = 0.5$, $\epsilon_{1L}= 1.0$, $\epsilon_{1U}= 2.0$, $\nu_L= -0.8$, $\nu_U= 0.8$, $\nu_{1L}= 0.3$, $\nu_{1U}= 0.4$, $\gamma_L= 0.5$, $\gamma_U= 1.0$. The chosen material parameters, including the auxetic phase ($\nu_L = -0.8$), were verified to satisfy the thermodynamic stability condition \eqref{eq:100} across the entire gradation range. On the left, the normalized transverse deflection $u_2$ predicted by the current 1D VAM-based theory (solid red line) is compared against the normalized mean transverse displacement $\langle w_2 \rangle$ obtained from the 3D FE benchmark (open circles). Additionally, the response from a "naive" rod theory (dashed black line)--which neglects the cross-sectional warping and the associated transverse energy correction $e_{\perp}$--is plotted. While the "naive" theory yields a significant error of approximately $20\%$ in predicting the maximum deflection, the current theory reduces this discrepancy to $2.48\%$. This stark contrast quantifies the essential physical effect of stiffening induced by the warping-related transverse constraints, which effectively increases the bending stiffness beyond the classical longitudinal fiber integration. To verify the theoretical error bounds, the relative $L_2$ error, defined by the ratio of squared differences as:
\begin{equation}
\label{L2error}
\| e \|_{L_2}^2 = \dfrac{\int_{0}^{L} | \langle w_2 \rangle - u_2 |^2 \dd{x}}{\int_{0}^{L} | \langle w_2 \rangle |^2 \dd{x}}
\end{equation}
is plotted against the rod slenderness $L/h$ on a log-log scale on the right of this Figure. The observed slope of approximately $-2$ provides robust numerical evidence for the asymptotic exactness of the theory, confirming that the modeling error of the displacement field is of the order $O(h/L)$.

\subsection{Extension to low-frequency vibrations}
The error estimation framework established for the static case generalizes naturally to low-frequency vibrations. Let $\check{\mathbf{u}}(x, t)$ denote the exact 3D elastodynamic solution for the anisotropic FG-rod, and $\mathbf{u}^*(x, t)$ be the reconstructed field derived from the dimensionally reduced 1D model. By linearity, the error field $\mathbf{w} = \check{\mathbf{u}} - \mathbf{u}^*$ satisfies the momentum balance equation:
\begin{equation}
\label{error}
\grad \vdot \boldsymbol{\sigma}(\mathbf{w}) + \mathbf{R} = \rho \ddot{\mathbf{w}},
\end{equation}
where $\mathbf{R}$ represents the dynamic residual, or the imbalance of the equations of motion induced by the reduction. Treating $\mathbf{R}$ as an effective body force, the rate of change of the error energy $\mathcal{E}_{err}$ is governed by:
\begin{equation}
\label{conserv}
\frac{d}{dt} \mathcal{E}_{err}(t) =\int_{\mathcal{V}} \mathbf{R} \cdot \dot{\mathbf{w}} \dd[3]{x} ,
\end{equation}
where the total error energy is defined as the sum of the kinetic and strain energy of the error field:
\begin{equation}
\label{errorE}
\mathcal{E}_{err}(t) = \frac{1}{2} \int_{\mathcal{V}} \rho \dot{\mathbf{w}}^2 \dd[3]{x} + \frac{1}{2} \int_{\mathcal{V}} \boldsymbol{\epsilon}(\mathbf{w}) \boldsymbol{:} \mathbf{C} \boldsymbol{:} \boldsymbol{\epsilon}(\mathbf{w}) \dd[3]{x}.
\end{equation}
Applying the Cauchy-Schwarz inequality to the right-hand side of Eq. \eqref{conserv} yields the bound:
\begin{equation}
\label{bound}
\int_{\mathcal{V}} \mathbf{R} \cdot \dot{\mathbf{w}} \dd[3]{x} \le \|\mathbf{R}\|_{L^2} \|\dot{\mathbf{w}}\|_{L^2}
\end{equation}
Observing that $\|\dot{\mathbf{w}}\|_{L^2} \le \sqrt{2 \mathcal{E}_{err} / \rho_{\min}}$, where $\rho_{\min}$ is the minimum density of the FG material, Eq. \eqref{conserv} is transformed into the following differential inequality:
\begin{equation}
\label{bound1}
\frac{d}{dt} \mathcal{E}_{err} \le \|\mathbf{R}\|_{L^2} \sqrt{\frac{2}{\rho_{min}}} \sqrt{\mathcal{E}_{err}}.
\end{equation}
Integration with respect to time leads to the energetic error bound:
\begin{equation}
\label{bound2}
\sqrt{\mathcal{E}_{err}(t)} \le \sqrt{\mathcal{E}_{err}(0)} + \frac{1}{\sqrt{2\rho_{min}}} \int_0^t \|\mathbf{R}(\tau)\|_{L^2} d\tau .
\end{equation}
The magnitude of $\mathcal{E}_{err}(t)$ is thus controlled by the residual $\mathbf{R}$, which can be decomposed into static and inertial components:
\begin{equation}
\label{decomp}
\mathbf{R} = \mathbf{R}_{stat} + \mathbf{R}_{inertial}.
\end{equation}
The smallness of $\mathbf{R}_{stat}$ is a consequence of the VAM-based static minimization discussed in Section 8.1. The inertial part $\mathbf{R}_{inertial}$ remains negligible in the low-frequency regime, where the characteristic wavelength is large compared to the cross-sectional dimensions, i.e., $h/(v\tau) \ll 1$. Consequently, the energetic consistency of the 1D model is preserved in the dynamic domain.
 
\section{Wave propagation in bi-layered transversely isotropic rod}

Although the theoretical framework for dynamic error estimation has been established, this section provides further corroboration by comparing the dispersion characteristics of the proposed 1D model with the 3D exact elasticity solution for a bi-layered, transversely isotropic cylinder \cite{nayfeh1996general}. The benchmark analytical solution is obtained by solving the transcendental dispersion relation derived from the 3D equations of motion, ensuring strict adherence to interface continuity and traction-free boundary conditions. Asymptotic expansion of this transcendental equation in the long-wave limit is expected to recover the dispersion relation of the 1D rod theory, thereby verifying its asymptotic exactness in the dynamic regime.

\subsection{1D effective properties and dispersion relation}

Let us first compute the effective mass and extension stiffness of our 1D theory for a straight bi-layered rod with a circular cross-section. We assume that the inner (core) layer occupies the region $r < R_1$, while the outer layer (cladding) occupies $R_1 < r < R_2$. Given that the cross-sectional problem decouples into anti-plane and plane-strain sub-problems, the axisymmetric plane-strain problem involves only the radial warping function $y_r(r)$, which is found by minimizing the following functional:
\begin{equation}
\label{eq:197}
\Phi_\perp =\frac{2\pi }{\pi R_2^2}\int_0^{R_2} \Bigl[ \frac{1}{2}\lambda (y_{r,r}+\frac{1}{r}y_r+2\nu \gamma)^2+\mu (y_{r,r}+\nu \gamma)^2+\mu (\frac{1}{r}y_r+\nu \gamma)^2\Bigr] r \dd{r},
\end{equation}
where $\lambda$, $\mu$, and $\nu$ are piecewise constant functions and $\gamma$ is the extension measure. The corresponding Euler--Lagrange equation is:
\begin{equation}
\label{eq:198}
y_r^{\prime \prime}+\frac{1}{r}y^\prime_r-\frac{1}{r^2}y_r=0,
\end{equation}
with the general solution:
\begin{equation}
\label{eq:199}
y_r(r) = 
\begin{cases}
 a_1r    & \text{for } 0 < r < R_1, \\
 a_2r + b_2\frac{1}{r}    & \text{for } R_1 < r < R_2.
\end{cases}
\end{equation}
To ensure a finite displacement at the origin, the singular term in the core is omitted. The coefficients $a_1, a_2,$ and $b_2$ are determined by enforcing the continuity of displacement and normal stress at the interface $r = R_1$:
\begin{align}
\label{eq:200}
&y_{r1}(R_1)=y_{r2}(R_1), \\
&[(\lambda _1+2\mu _1) y_{r1}^\prime +\lambda _1\frac{1}{r}y_{r1}+2(\lambda _1+\mu_1)\nu_1 \gamma]\Bigl|_{r=R_1} \notag
\\
&=[(\lambda _2+2\mu _2) y_{r2}^\prime +\lambda _2\frac{1}{r}y_{r2}+2(\lambda _2+\mu_2)\nu _2 \gamma]\Bigl|_{r=R_1}, \label{eq:201}
\end{align} 
and the traction-free condition at the outer surface $r = R_2$:
\begin{equation}
\label{eq:202}
[(\lambda _2+2\mu _2) y_{r2}^\prime +\lambda _2\frac{1}{r}y_{r2}+2(\lambda _2+\mu_2)\nu_2 \gamma]_{r=R_2}=0.
\end{equation}
Solving this system yields:
\begin{align}
a_1 &= -\frac{\kappa_1 \kappa_2 \nu_1 + \mu_2 [f \kappa_1 \nu_1 + (1-f) \kappa_2 \nu_2]}{\kappa_1 \kappa_2 + \mu_2 [f \kappa_1 + (1-f) \kappa_2]} \gamma, \label{eq:203} \\
a_2 &= -\frac{\kappa_1 \kappa_2 \nu_2 + \mu_2 [f \kappa_1 \nu_1 + (1-f) \kappa_2 \nu_2]}{\kappa_1 \kappa_2 + \mu_2 [f \kappa_1 + (1-f) \kappa_2]} \gamma, \label{eq:204} \\
b_2 &= R_1^2 \frac{\kappa_1 \kappa_2 (\nu_2 - \nu_1)}{\kappa_1 \kappa_2 + \mu_2 [f \kappa_1 + (1-f) \kappa_2]} \gamma, \label{eq:205}
\end{align}
where $\kappa_i = \lambda_i + \mu_i$ and $f = (R_1/R_2)^2$ is the core volume fraction. Substituting these into the functional, the minimum energy is:
\begin{equation}
\label{eq:206}
\Phi_\perp = \frac{2 f (1-f) \mu_2 \kappa_1 \kappa_2 (\nu_1 - \nu_2)^2}{\kappa_1 \kappa_2 + \mu_2 [f \kappa_1 + (1-f) \kappa_2]} \gamma^2.
\end{equation}
Comparing this to $\Phi_\perp = \frac{1}{2}E_\perp \gamma^2$, the transverse correction to the stiffness is:
\begin{equation}
\label{eq:207}
E_\perp=\frac{4 f (1-f) \mu_2 \kappa_1 \kappa_2 (\nu_1 - \nu_2)^2}{\kappa_1 \kappa_2 + \mu_2 [f \kappa_1 + (1-f) \kappa_2]}.
\end{equation}
Combined with the effective properties $\bar{\rho} = \rho_1 f + \rho_2 (1-f)$ and $E_\parallel = E_1 f + E_2(1-f)$, the 1D governing equation is $\bar{\rho}\ddot{u} = \bar{E}u^{\prime \prime}$, where $\bar{E} = E_\parallel + E_\perp$. For a longitudinal harmonic wave $u e^{i(kx-\omega t)}$, the 1D dispersion relation is:
\begin{equation}
\label{eq:210}
\bar{\rho} \omega^2 = \bar{E}k^2.
\end{equation}

\subsection{Comparison with the 3D asymptotic limit}

In 3D elasticity, axisymmetric longitudinal waves are characterized by radial ($w_r$) and axial ($w$) displacements. Assuming harmonic forms $w_r = W_r(r)e^{i(kx-\omega t)}$ and $w = W(r)e^{i(kx-\omega t)}$, the governing equations reduce to Bessel-type ODEs:
\begin{align}
(\lambda + 2\mu) \Bigl( W_r'' + \frac{W_r'}{r} - \frac{W_r}{r^2}  \Bigr) + (\rho \omega^2 - G k^2) W_r - i k (\lambda_{31} + G) W' &= 0, \label{eq:213} \\
G \Bigl( W'' + \frac{W'}{r}  \Bigr) + (\rho \omega^2 - E k^2) W - i k (\lambda_{31} + G) \Bigl( W_r' + \frac{W_r}{r} \Bigr) &= 0. \label{eq:214}
\end{align}
The solutions in the layers are expressed using Bessel functions $J_n$ and $Y_n$:
\begin{align}
W_{r1}(r) &= A_1J_1(q_{11}r)+C_1J_1(q_{12}r), \label{eq:215} \\ 
W_1(r) &= ik [\alpha_{11} A_1J_0(q_{11}r)+\alpha_{12} C_1J_0(q_{12}r)], \label{eq:216}
\end{align}
with similar forms for $W_{r2}$ and $W_2$ in the cladding, incorporating $Y_0$ and $Y_1$. The radial wavenumbers $q_{ij}$ are roots of the Christoffel equation 
\begin{equation}
a_i(q^2)^2 + b_i(q^2) + c_i = 0,
\label{eq:219} 
\end{equation}
where
\begin{align}
&a_i = (\lambda_i + 2\mu_i)G_i, \label{eq:220}\\
&b_i = (\lambda_i + 2\mu_i)(E_i k^2 - \rho_i \omega^2) + G_i(G_i k^2 - \rho_i \omega^2) - (\lambda_{31}^{(i)} + G_i)^2 k^2, \label{eq:221}\\
&c_i = (G_i k^2 - \rho_i \omega^2)(E_i k^2 - \rho_i \omega^2), \label{eq:222}
\end{align}
and $\alpha_{ij}$ denotes the amplitude ratio
\begin{equation}
\label{eq:223}
\alpha_{ij} = \frac{(\lambda_i + 2\mu_i) q_{ij}^2 + G_i k^2 - \rho_i \omega^2}{k q_{ij} (\lambda_{31}^{(i)} + G_i)}.
\end{equation}

The transcendental dispersion relation is obtained by setting the determinant of the $6 \times 6$ coefficient matrix--derived from the continuity of displacements and stresses at the interface and traction-free conditions at the outer boundary--to zero (see \cite{nayfeh1996general}). In the long-wave limit ($k \to 0, \omega \to 0$), the Bessel functions are replaced by their power series approximations. The 3D system decouples at leading order, and the longitudinal mode assumes a constant axial displacement $W \approx W_0$. This simplifies the boundary-value problem to a $3 \times 3$ system for the radial constants:
\begin{equation}
\begin{pmatrix} 
R_1 & -R_1 & -1/R_1 \\
2\kappa_1 & -2\kappa_2 & 2\mu_2/R_1^2 \\
0 & 2\kappa_2 & -2\mu_2/R_2^2 
\end{pmatrix} 
\begin{pmatrix} \mathcal{A}_1 \\ \mathcal{A}_2 \\ \mathcal{B}_2 \end{pmatrix} = 
\begin{pmatrix} 0 \\ (\lambda_{31}^{(2)} - \lambda_{31}^{(1)}) \gamma \\ -\lambda_{31}^{(2)} \gamma \end{pmatrix}. \label{eq:231}
\end{equation}
Solving this gives coefficients $\mathcal{A}_1, \mathcal{A}_2,$ and $\mathcal{B}_2$ that coincide exactly with the warping parameters $a_1, a_2,$ and $b_2$ from the 1D theory. By evaluating the global momentum balance $\langle \rho \ddot{w} \rangle = \langle \sigma_{xx} \rangle_{,x}$ and substituting the recovered 3D stresses, we obtain:
\begin{equation}
\label{eq:237}
\bar{\rho} \omega^2 = \bar{E}k^2.
\end{equation}
The identical nature of \eqref{eq:237} and \eqref{eq:210} confirms the asymptotic exactness of the proposed 1D rod theory in the dynamic regime proved in subsection 8.3.

\section{Conclusion}

This study has presented a comprehensive VAM-based framework for functionally graded anisotropic rods. The primary contribution is the establishment of an asymptotically exact dimension reduction procedure that avoids ad hoc kinematic assumptions. A key discovery is that classical 'naive' beam theory, based on simple longitudinal fibre integration, can incur errors up to $20\%$ in predicting deflections, primarily due to their failure to account for transverse constraints and Poisson’s ratio mismatch. In contrast, the current VAM-based cross-sectional analysis identifies a significant stiffening effect induced by the warping-related transverse interactions. By incorporating the energetic correction $e_{\perp}$, the proposed theory reduces the modeling error to approximately $2\%$, providing a high-fidelity representation of the true 3D physics. The key contributions of this research include:
\begin{enumerate}
\item Dual-Bounding Framework: The formulation of a dual-bounding approach that provides rigorous upper and lower bounds for the effective 1D stiffness coefficients, ensuring numerical stability and reliability.
\item Proof of Asymptotic Exactness: A formal proof that the theory is asymptotically exact in both statics and dynamics, as verified through the Prager-Synge identity and the recovery of exact 3D dispersion relations in the long-wave limit.
\item Systematic 3D Restoration: A robust procedure for reconstructing the local 3D stress and strain fields, which accurately captures the complex effects of material gradation and auxeticity.
\end{enumerate}
The developed framework serves as a powerful tool for the design of high-performance acoustic waveguides, thin-walled structural components, and anisotropic actuators. Future work will extend this variational approach to account for multi-physical coupling in piezoelectric FG materials and non-linear large-deflection regimes.

\appendix
\section{$\vb{c}$-matrix and $\vb{f}$- and $\vb{g}$-vectors of the direct and dual problems}  

For the direct problem, we set $\mathbf{a}=0$ and $\mathbf{u} = (y_1, y_2, y)^T$. By comparing the first variation of the transverse energy \eqref{eq:41} with the standard form \eqref{eq:74}, the $6 \times 6$ coefficient matrix $\mathbf{c}$ is assembled from the 3D stiffness components as follows:
\begin{equation}
\label{eq:77}
\mathbf{c} = \begin{pmatrix}
c_{1111} & c_{1112} & c_{1121} & c_{1122} & c_{1131} & c_{1132} \\
c_{1211} & c_{1212} & c_{1221} & c_{1222} & c_{1231} & c_{1232} \\
c_{2111} & c_{2112} & c_{2121} & c_{2122} & c_{2131} & c_{2132} \\
c_{2211} & c_{2212} & c_{2221} & c_{2222} & c_{2231} & c_{2232} \\
c_{3111} & c_{3112} & c_{3121} & c_{3122} & c_{3131} & c_{3132} \\
c_{3211} & c_{3212} & c_{3221} & c_{3222} & c_{3231} & c_{3232}
\end{pmatrix}.
\end{equation}
The elements of corresponding source vector $\mathbf{f} = (f_1, f_2, f)^T$ and boundary vector $\mathbf{g} = (g_1, g_2, g)^T$ are:
\begin{align}
f_1&=(c_{1133}(\gamma +h\Omega_\sigma \zeta_\sigma ))_{,1}+(c_{1233}(\gamma +h\Omega_\sigma \zeta_\sigma ))_{,2}-h\Omega (c_{1131}\zeta_2-c_{1132}\zeta_1)_{,1} \notag
\\
&-h\Omega (c_{1231}\zeta_2-c_{1232}\zeta_1)_{,2}+((c_{1131}r_1+c_{1132}r_2)(\gamma +h\Omega_\sigma \zeta_\sigma ))_{,1} \notag 
\\
&+((c_{1231}r_1+c_{1232}r_2)(\gamma +h\Omega_\sigma \zeta_\sigma ))_{,2}, \label{eq:78} 
\end{align}
\begin{align}
f_2&=(c_{2133}(\gamma +h\Omega_\sigma \zeta_\sigma ))_{,1}+(c_{2233}(\gamma +h\Omega_\sigma \zeta_\sigma ))_{,2}-h\Omega (c_{2131}\zeta_2-c_{2132} \zeta_1)_{,1} \notag
\\
&-h\Omega (c_{2231}\zeta_2-c_{2232}\zeta_1)_{,2}+((c_{2131}r_1+c_{2132}r_2)(\gamma +h\Omega_\sigma \zeta_\sigma ))_{,1}  \notag
\\
&+((c_{2231}r_1+c_{2232}r_2)(\gamma +h\Omega_\sigma \zeta_\sigma ))_{,2}, \label{eq:79}
\end{align}
\begin{align}
f_3&=-h\Omega (c_{3131}\zeta_2-c_{3132}\zeta_1)_{,1}-h\Omega (c_{3231}\zeta_2-c_{3232}\zeta_1)_{,2} \notag
\\
&+((c_{3131}r_1+c_{3132}r_2)(\gamma +h\Omega_\sigma \zeta_\sigma ))_{,1}
+((c_{3231}r_1+c_{3232}r_2)(\gamma +h\Omega_\sigma \zeta_\sigma ))_{,2} \notag
\\
&+((c_{3111}r_{11}+c_{3122}r_{22}+2c_{3112}r_{12})(\gamma +h\Omega_\sigma \zeta_\sigma ))_{,1} \notag
\\
&+((c_{3211}r_{11}+c_{3222}r_{22}+2c_{3212}r_{12})(\gamma +h\Omega_\sigma \zeta_\sigma ))_{,2}, \label{eq:80} 
\end{align}
\begin{align}
g_1&=c_{1133}(\gamma +h\Omega_\sigma \zeta_\sigma )n_{1}+c_{1233}(\gamma +h\Omega_\sigma \zeta_\sigma )n_{2}-h\Omega (c_{1131}\zeta_2-c_{1132}\zeta_1)n_{1} \notag
\\
&-h\Omega (c_{1231}\zeta_2-c_{1232}\zeta_1)n_{2}+(c_{1131}r_1+c_{1132}r_2)(\gamma +h\Omega_\sigma \zeta_\sigma )n_{1} \notag 
\\
&+(c_{1231}r_1+c_{1232}r_2)(\gamma +h\Omega_\sigma \zeta_\sigma )n_{2}, \label{eq:81}
\end{align}
\begin{align}
g_2&=c_{2133}(\gamma +h\Omega_\sigma \zeta_\sigma )n_{1}+c_{2233}(\gamma +h\Omega_\sigma \zeta_\sigma )n_{2}-h\Omega (c_{2131}\zeta_2-c_{2132} \zeta_1)n_{1} \notag
\\
&-h\Omega (c_{2231}\zeta_2-c_{2232}\zeta_1)n_{2}+(c_{2131}r_1+c_{2132}r_2)(\gamma +h\Omega_\sigma \zeta_\sigma )n_{1}  \notag
\\
&+(c_{2231}r_1+c_{2232}r_2)(\gamma +h\Omega_\sigma \zeta_\sigma )n_{2}, \label{eq:82}
\end{align}
\begin{align}
g_3&=-h\Omega (c_{3131}\zeta_2-c_{3132}\zeta_1)n_{1}-h\Omega (c_{3231}\zeta_2-c_{3232}\zeta_1)n_{2} \notag
\\
&+(c_{3131}r_1+c_{3132}r_2)(\gamma +h\Omega_\sigma \zeta_\sigma )n_{1}
+(c_{3231}r_1+c_{3232}r_2)(\gamma +h\Omega_\sigma \zeta_\sigma )n_{2} \notag
\\
&+(c_{3111}r_{11}+c_{3122}r_{22}+2c_{3112}r_{12})(\gamma +h\Omega_\sigma \zeta_\sigma )n_{1} \notag
\\
&+(c_{3211}r_{11}+c_{3222}r_{22}+2c_{3212}r_{12})(\gamma +h\Omega_\sigma \zeta_\sigma )n_{2}, \label{eq:83} 
\end{align}
where $\Omega_\sigma \zeta_\sigma =\Omega_1 \zeta_1+\Omega_2 \zeta_2 $. 

For the dual problem, we seek the stress functions $\mathbf{u} = (\psi_1, \psi_2, \psi)^T$. The dual $\mathbf{c}$-matrix is derived from the 2D compliance tensor $\bar{s}$ and the penalty parameter $\theta$ used to enforce the symmetry of the stress tensor:
\begin{equation}
\label{eq:92}
\mathbf{c}^* = \begin{pmatrix}
\bar{s}_{2222} + \theta & -\bar{s}_{2212} & -\bar{s}_{2221} & \bar{s}_{2211} - \theta & \bar{s}_{222} & -\bar{s}_{221} \\
-\bar{s}_{1222} & \bar{s}_{1212} + \theta & \bar{s}_{1221} - \theta & -\bar{s}_{1211} & -\bar{s}_{122} & \bar{s}_{121} \\
-\bar{s}_{2122} & \bar{s}_{2112} - \theta & \bar{s}_{2121} + \theta & -\bar{s}_{2111} & -\bar{s}_{212} & \bar{s}_{211} \\
\bar{s}_{1122} - \theta & -\bar{s}_{1112} & -\bar{s}_{1121} & \bar{s}_{1111} + \theta & \bar{s}_{112} & -\bar{s}_{111} \\
\bar{s}_{222} & -\bar{s}_{212} & -\bar{s}_{122} & \bar{s}_{112} & \bar{s}_{22} & -\bar{s}_{21} \\
-\bar{s}_{221} & \bar{s}_{211} & \bar{s}_{121} & -\bar{s}_{111} & -\bar{s}_{12} & \bar{s}_{11}
\end{pmatrix}.
\end{equation}
The components of $\vb{f}$-vector read:
\begin{align}
f_1&=-(r_{22}(\gamma +h\Omega_\sigma \zeta_\sigma ))_{,1}+(r_{21}(\gamma +h\Omega_\sigma \zeta_\sigma ))_{,2}, \label{eq:93} 
\\
f_2&=(r_{12}(\gamma +h\Omega_\sigma \zeta_\sigma ))_{,1}-(r_{11}(\gamma +h\Omega_\sigma \zeta_\sigma ))_{,2}, \label{eq:94}
\\
f_3&=2h\Omega +(r_{2}(\gamma +h\Omega_\sigma \zeta_\sigma ))_{,1}-(r_{1}(\gamma +h\Omega_\sigma \zeta_\sigma ))_{,2}. \label{eq:95} 
\end{align}
The $\vb{g}$-vector is not required, as the homogeneous Dirichlet boundary conditions are imposed on $\psi_\alpha$ and $\psi$.

\section*{Declaration of generative AI and AI-assisted technologies in the writing process}

During the preparation of this work the authors used Google Gemini in order to improve language and readability. After using this tool/service, the authors reviewed and edited the content as needed and take full responsibility for the content of the publication.

\bibliographystyle{elsarticle-num}
\bibliography{references}

\end{document}